%% file: journal.tex
\pdfoutput=1
\documentclass[12pt, draftclsnofoot, onecolumn]{IEEEtran}
\ifCLASSINFOpdf
  \usepackage[pdftex]{graphicx}
  \graphicspath{{.}}
  \DeclareGraphicsExtensions{.pdf,.jpeg,.png}
\else
  \usepackage[dvips]{graphicx}
  \graphicspath{{.}}
  \DeclareGraphicsExtensions{.pdf}
\fi

\usepackage{graphics}
\usepackage{cite}
\usepackage{acronym}
\usepackage{units}
\usepackage{amsmath}
\usepackage{amsfonts}
\usepackage{amssymb}
\usepackage[center]{subfigure}
\usepackage{multirow}
\usepackage{booktabs}
\usepackage{array}

\IEEEoverridecommandlockouts
\ifCLASSINFOpdf
\else
\fi

\usepackage[draft,nomargin,marginclue,footnote,silent]{fixme}

\usepackage{prettyref}
\newrefformat{chap}{Chapter~\ref{#1}}
\newrefformat{sec}{Sec.~\ref{#1}}
\newrefformat{subsec}{Sec.~\ref{#1}}
\newrefformat{eq}{Eq.~(\ref{#1})}
\newrefformat{fig}{Fig.~\ref{#1}}
\newrefformat{ex}{Example~\ref{#1}}
\newrefformat{list}{List~\ref{#1}}
\newrefformat{tab}{Table~\ref{#1}}
\newrefformat{enum}{Case~(\ref{#1}.)}
\newrefformat{def}{Definition~\ref{#1}}
\newrefformat{le}{Lemma~\ref{#1}}
\newrefformat{th}{Theorem~\ref{#1}}
\newcommand{\pref}[1]{\prettyref{#1}}

\newcommand{\eg}{e.\,g., }

\newcommand{\ie}{i.\,e., }

\newcommand{\cf}{cf.~}

\newcommand{\direct}{\textsc{Direct}}
\newcommand{\dname}{\textsc{Best-Relay}}
\newcommand{\cname}{\textsc{Best-Antenna}}

\def\scalecaption{}
\newcommand{\mycap}{REPLACE with renewcommand mycap}

\newcommand{\doublefig}[7]{
\def\myhspace{-1.25em}
\def\mywidth{#7}
\def\myvspace{-10pt}
\def\myhalfvspace{-0.3em}
\begin{figure}[#6]
 \centering
  \subfigure[#2]{
    \label{#1-1}
     \hspace{-1em}
    \includegraphics[width=\mywidth\columnwidth]{#3}}%
         \vspace{\myhalfvspace}
\hspace{\myhspace}
  \subfigure[#4]{
    \label{#1-2}
     \hspace{-1em}
    \includegraphics[width=\mywidth\columnwidth]{#5}}%
 \caption{\scalecaption \mycap  \label{#1}}
     \vspace{-10pt}
\end{figure}
}

\acrodef{ACK}{Acknowledgement}
\acrodef{ARQ}{Automatic Repeat reQuest}
\acrodef{AWGN}{Additive White Gaussian Noise}
\acrodef{M2M}{Machine-to-Machine Communications}
\acrodef{COTS}{Commercial Off-The-Shelf}
\acrodef{IEEE}{Institute of Electrical and Electronics Engineers}
\acrodef{CSMA/CA}{Carrier Sense Multiple Access with Collision Avoidance}
\acrodef{MAC}{Medium Access Control}
\acrodef{STBC}{Space Time Block Code}
\acrodef{CSI}{Channel State Information}
\acrodef{AP}{Access Point}
\acrodef{TDMA}{Time Division Multiple Access}
\acrodef{Tx}{Transmitting Terminal}
\acrodef{Rx}{Receiving Terminal}
\acrodef{BP}{Beacon Period}
\acrodef{TP}{Transmission Period}
\acrodef{CP}{CSI Period}
\acrodef{PDF}{Probability Density Function}
\acrodef{CDF}{Cumulative Distribution Function}
\acrodef{SNR}{Signal to Noise Ratio}
\acrodef{PER}{Packet Error Rate}
\acrodef{MER}{Message Error Rate}
\acrodef{BER}{Bit Error Rate}
\acrodef{QoS}{Quality-of-Service}
\acrodef{PDR}{packet delivery ratio}
\acrodef{IBL}{Infinite Blocklength}
\acrodef{FBL}{Finite Blocklength}

\begin{document}
\title{Finite Blocklength Performance of\\Multi-Terminal  Wireless Industrial Networks}

\author{Yulin~Hu, Martin~Serror, Klaus~Wehrle, and James~Gross }
\maketitle

\begin{abstract}
This work focuses on the performance of multi-terminal  wireless industrial networks, where the transmissions of all terminals are required to be scheduled within a tight deadline.
The transmissions thus share a fixed amount of resources, i.e., symbols, while facing short blocklengths due to the low-latency requirement. 
We investigate two distinct relaying strategies, namely best relay selection among the participating terminals and best antenna selection at the access point of the network.
In both schemes, we incorporate the cost of acquiring instantaneous \ac{CSI} at the access point within the transmission deadline. 
An error probability model is developed under the finite blocklength regime to provide accurate performance results.
As a reference, this model is compared to the corresponding infinite bocklength error model.
Both analytical models are validated by simulation.
We show that the average \ac{PER} over all terminals is convex in the target error probability at each single link. 
Moreover, we find that: 
(i)~The reliability behavior is different for the two strategies, while the limiting factors are both finite blocklengths and overhead of acquiring \ac{CSI}.  
(ii)~With the same order of diversity, best antenna selection is more reliable than best relay selection. 
(iii)~The average \ac{PER} is increasing in the number of participating terminals unless the terminals also act as relay candidates.  
In particular, if each participating terminal is a candidate for best relay selection, the \ac{PER} is convex in the number of terminals. 
\end{abstract}


% \fi
\IEEEpeerreviewmaketitle

\acresetall

\input{01_intro.tex}

\input{04_system_model.tex}

\input{05_error_probability.tex}

\input{06_outage_IBL_regime.tex}

\input{07_evaluation.tex}

\input{08_conclusion.tex}
\appendices
\input{09_proof_2}

% or


%%\appendices
%%\section{Proof of the First Zonklar Equation}
%%Appendix one text goes here.

%%\section{}
%%Appendix two text goes here.

 
%\section*{Acknowledgment}

\ifCLASSOPTIONcaptionsoff
  \newpage
\fi



%\bibliographystyle{IEEEtran}
%\begin{thebibliography}{1}
%\end{thebibliography}

\bibliographystyle{IEEEtran}
\bibliography{IEEEfull,journal}


%%\begin{IEEEbiography}{Michael Shell}
%%Biography text here.
%%\end{IEEEbiography}

%%\begin{IEEEbiographynophoto}{John Doe}
%%Biography text here.
%%\end{IEEEbiographynophoto}

%\newpage

%%\begin{IEEEbiographynophoto}{Jane Doe}
%%Biography text here.
%%\end{IEEEbiographynophoto}


%\vfill


\end{document}

%% file: 01_intro.tex
\section{Introduction}
\label{sec:intro}
 
%\IEEEPARstart{T}
The proliferation of \ac{M2M} in home, business and industrial environments entails new requirements towards wireless communications.
Besides optimizing spectral efficiency, future wireless communication standards, such as 5G, must support ultra-low latency communication at predictable high reliabilities~\cite{ABC+14}.
In industrial automation, for example, safety- and mission-critical applications have stringent requirements regarding \ac{QoS}, which are currently not met by existing wireless standards~\cite{FWB+14}.
Anticipated target bounds for reliability and latency are typically around \unit[$1-10^{-9}$]{\ac{PDR}} and \unit[1]{ms}, respectively~\cite{Neum07}.
Thus, efficient ways must be explored to increase the communication reliability of wireless networks while complying to the ultra-low latency bound.
More importantly, accurate performance models of these schemes must be proposed to allow for sound design decisions of such systems.

It is well known that reliability is increased by exploiting diversity in time, frequency and/or space.
It has been shown that when operating on very short time scales, spatial diversity is especially beneficial for increasing the communication reliability, making use of additional uncorrelated transmission paths~\cite{DASC04}. 
Moreover, cooperative diversity, a special form of spatial diversity, allows leveraging distributed resources of overhearing terminals.
This is especially useful when the considered terminals have hardware constraints, \eg when they are limited to a single transceiver antenna, allowing the terminals to perform relaying or even form a virtual antenna array.
It is known that cooperative diversity, \eg cooperative \ac{ARQ}, reduces the outage probability by several orders of magnitude in wireless communications~\cite{LTWo04}.
A common approach to further enhance the reliability in cooperative networks is to increase the number of cooperation relays. 
Laneman et al.~\cite{LTWo04} show that full diversity order in the number of cooperating terminals can be achieved.
In~\cite{BKRL06, BlLi06}, a simple scheme is proposed for selecting the ``best'' relay out of several potential relays based on end-to-end instantaneous \ac{CSI}.  
It is shown that this approach achieves the same performance as more complex space-time coding.
A closed-form expression for the outage probability is provided in~\cite{IkAh10}.
The authors of~\cite{ChSe16} investigate the impact on the transmission delay when using relaying compared to direct transmissions, \ie under which conditions relaying improves the end-to-end transmission delay.
A latency analysis is derived under the assumption of a Gaussian channel, not including the effects of a fading channel.
In~\cite{SSR+15}, the authors address high reliable, low latency wireless networks by proposing a cooperative approach in which nodes simultaneously relay messages to reduce the outage probability.
Their approach is evaluated assuming Rayleigh fading and infinite blocklengths.
The results show that the transmission reliability increases with the number of participating nodes, even for a low cycle time of \unit[2]{ms}.
Likewise, in~\cite{DoGr15} a wireless real-time protocol is presented that can achieve latencies within a few milliseconds while providing extremely high reliabilities.
This is achieved through cooperative ARQ while the authors even demonstrated these results through experimental results of a prototype.
Comparably, we showed in previous work~\cite{SDWG15} that cooperative \ac{ARQ} can be effectively integrated into a multi-terminal \ac{TDMA} system with a stringent time deadline.

However, typically these studies are based idealistic assumptions, namely not considering overhead for acquiring \ac{CSI} as well as arbitrarily reliable communication at Shannon's channel capacity which strictly speaking can only be achieved by coding with infinite blocklengths.
Unfortunately, both of these assumptions are too optimistic in practice.  
Wireless networks are likely to be comprised of multiple terminals with a significant number of links between the terminals.
Hence, the overhead of acquiring \ac{CSI} of these links is considerable, increasing with each additional terminal. 
More importantly, low-latency bounds in combination with more and more terminals sharing a fixed amount of symbols lead to short blocklengths, which are known to have a different error performance even if communicating below the Shannon capacity which is based on the infinite blocklength assumption.
In~\cite{Verdu_2010} it was shown that the performance difference between infinite blocklength (i.e. Shannon capacity) and finite blocklength is considerable and increases for shorter and shorter blocklengths. 
This indicates that the results of existing research, based on outage capacity models stemming from the infinite blocklength assumption, are inaccurate, \eg \cite{DASC04, SHD+16}. 
The effects on the performance of single-terminal relaying under the finite blocklength assumption were extensively investigated in~\cite{Hu_2015, Hu_letter_2015, Hu_TWC_2016}.
Nevertheless, there is a lack of performance evaluations of multi-terminal systems, where transmission resources are shared and instantaneous \ac{CSI} must be acquired while a larger number of terminals leads on the other hand to a higher diversity degree.

In this work, we investigate whether high reliability can be achieved with cooperative relaying in latency-constrained, multi-terminal wireless networks under realistic assumptions regarding blocklengths and \ac{CSI} overhead.
In our analysis, we thus focus on the effects of finite blocklengths and on the overhead of acquiring instantaneous \ac{CSI} on the communication reliability.
A growing number of participants in a cooperative network potentially increases the diversity degree while the blocklengths for the individual transmissions decrease.
Moreover, as more links must be considered for the relaying paths, the overhead for the collection of \ac{CSI} increases as well, which additionally reduces the available transmission blocklengths.
The fundamental questions addressed in this paper thus are: 
How reliable can such a wireless network get at a given (low) target latency?
Which design decisions should be considered to achieve the anticipated reliability? 

We introduce in the following two system variants which both exploit cooperative transmission paired with perfect CSI.
Our system model accounts in these settings on the one hand for the overhead of operating such systems, while on the other hand we then derive bounds on the reliability of the system based on outage capacity and finite blocklength error models. 
Based on these models we provide the following novel contributions:
\begin{itemize}
  \item We characterize the error performance of cooperative multi-terminal wireless systems under the \ac{FBL} regime and show in particular that the error performance of a single, tracked terminal, as well as the overall multi-terminal error performance is convex in the decoding error probability with which the individual links are operated.
  \item We provide an error performance comparison of the cooperative systems under the \ac{IBL} as well as the \ac{FBL} regime, and can show that the impact due to \ac{FBL} modeling is significant, leading to a different qualitative and quantitative behavior of the investigated systems. This is relevant for the design of such systems, as the results clearly show that any low latency design that does not take \ac{FBL} effects into account is likely to result in different, erroneous design decisions. 
  \item Numerically we can show that as long as the cooperative diversity degree increases while also the system load increases, the overall error performance of the system improves \textit{despite} accounting for the overhead and the \ac{FBL} effects. 
\end{itemize}

The remainder of this paper is structured as follows.
The system model assumptions are presented in \pref{sec:system}.
In \pref{sec:finite}, we derive the \ac{PER} under the \ac{FBL} regime; the key performance indicator of the considered system.
In \pref{sec:infinite}, we discuss the \ac{PER} in the \ac{IBL} regime, this will serve as a reference for the effects of short blocklengths on the system performance.
A validation and numerical evaluation of the introduced models is included in \pref{sec:eval}.
A conclusion of this paper is provided in \pref{sec:conclusion}.

%% file: 04_system_model.tex
\section{System Model and Problem Statement}
\label{sec:system}
In this section, we first give a general description of the system model as well as the two considered variants. 
Afterward, we introduce the considered error models and the overhead models. 
Then, we propose a cost model to account for the effects of periodically collecting instantaneous \ac{CSI}. 
Finally, we formulate the problem statement that we address in the further course of this paper.
  
\subsection{General System Model}
We consider a wireless network for ultra-reliable and low-latency communication.
The network consists of an \ac{AP} and $N$ associated terminals, which are all in communication range of each other, \ie terminals can   directly send packets to each other and also overhear the transmissions from other terminals.
The considered transmission medium is assumed to be a flat radio channel, operating over a given bandwidth $B$.
Transmissions are mainly affected by fading, which we model by a Rayleigh-distributed block-fading channel.
The instantaneous quality of a link is characterized by the \ac{SNR}.
We denote by $\gamma_{i,j}$, with $i,j=0,1,\ldots, N \land i \ne j$, the \ac{SNR} of the link from terminal $i$ to terminal $j$, where $i=0$ or $j=0$ indicates the link from or to the \ac{AP}. 
Furthermore, we assume all links to be reciprocal, \ie $\gamma_{i,j} = \gamma_{j,i}$.
Due to the varying nature of the wireless channel, $\gamma_{i,j}$ varies over time around the average value $\overline{\gamma}_{i,j}$. 
In particular, $\gamma_{i,j} =  z \overline{\gamma}_{i,j}$, where $z$ is the channel fading gain with \ac{PDF}:
\begin{IEEEeqnarray}{RCL}
\label{eq:gamma_distribution}
f \left( z  \right) & = &  \exp \left( - z \right) \quad .
\end{IEEEeqnarray}

To realize guaranteed access to the shared communication medium, we consider a \ac{TDMA} system where the \ac{AP} centrally assigns time slots to the associated terminals.
In general, terminals are assumed to have limited hardware resources, \ie only one transmission antenna due to space and cost constraints while the \ac{AP} could be equipped with multiple antennas.
Particularly, for a system variant with multiple antennas at the \ac{AP}, we assume that the average \ac{SNR} of the links between a terminal~$i$ and the different antennas of the \ac{AP} are homogeneous and correspond to~$\overline{\gamma}_{i,0}$ and accordingly~$\overline{\gamma}_{0,i}$. 

A central requirement  of the system is to ensure high transmission reliability within a fixed latency bound.
In other words, for each of the $N$ associated terminals, we want to guarantee a reliable transmission, \ie below a certain \ac{PER}, of a packet of size $D$ (in bits), within a cycle time $T_\mathrm{cyc}$.  
We are interested in the performance of cooperative transmission schemes, \ie a packet from a \ac{Tx} to a \ac{Rx} may be either transmitted directly or it is relayed via a third cooperating terminal depending on the link conditions.
To reduce the packet error probability, a transmission path between \ac{Tx} and \ac{Rx} should be selected providing the highest reliability in terms of link conditions.
Consequently, the \ac{AP}, which is responsible for the scheduling, periodically acquires instantaneous \ac{CSI} about the links in the network and schedules transmission paths accordingly.

\begin{figure}
\centering
\includegraphics[width=0.6\columnwidth]{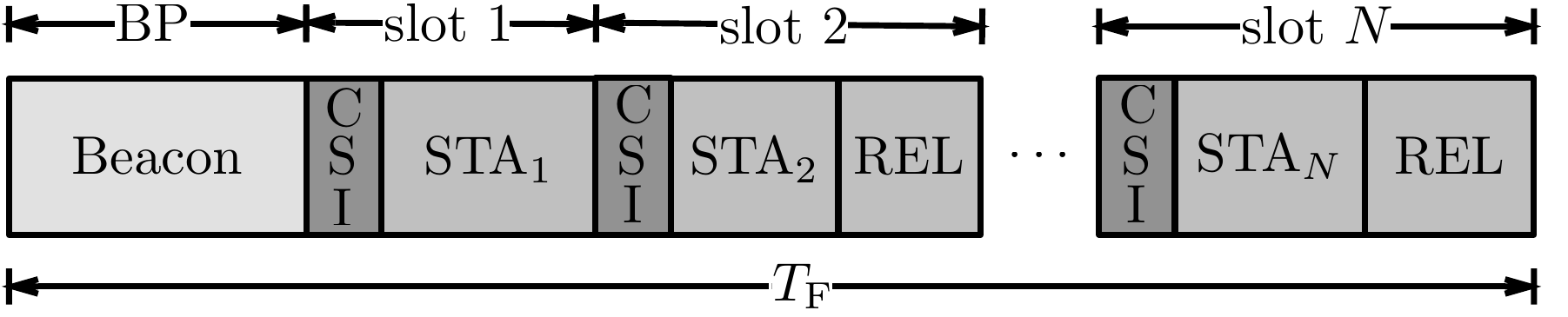}
\caption{Example of the \acs{TDMA} frame structure for the two relaying strategies. After the \acs{BP}, the frame is divided into $N$ slots, corresponding to the $N$ packets that need to be transmitted. Within a slot, a packet is either transmitted directly to the receiver or via a relay. Instantaneous \acs{CSI} is piggybacked within the transmissions to the overhearing \acs{AP}.}
\label{fig:frame-tdma}
\end{figure}

The considered \ac{TDMA} frame is depicted in \pref{fig:frame-tdma}.
It consists of a \ac{BP} and a \ac{TP}.
In the \ac{BP}, the \ac{AP} sends a packet, which includes a transmission schedule and serves as a synchronization reference for the associated terminals.
The \ac{TP} has a fixed total length of $S$ symbols.
It is further divided into $N$ slots with arbitrary blocklengths, each reserved for one of the associated terminals and determined by a scheduler.
Each blocklength individually depends on the considered link qualities and on whether a direct or an indirect transmission path was selected by the \ac{AP}.
At the beginning of each slot, a certain amount of time is reserved for the estimation of instantaneous \ac{CSI} of the links. 
Therefore, \ac{Tx} transmits a reference signal to \ac{Rx}, which enables \ac{Rx} to estimate the current link quality.
This information must be then conveyed to the \ac{AP}, which centrally collects \ac{CSI} for the scheduling decisions.
To reduce the time overhead, a terminal piggybacks the most recent \ac{CSI} values in a subsequent transmission, which is overhead by the \ac{AP}.
Details on the cost of acquiring \ac{CSI} are provided in \pref{sec:csi-acquisition}.
As we are interested in the performance of cooperative transmission schemes, we now sketch two different organizations of a cooperative system which we use in the following as base for our analysis. 
We refer to them as \cname{} and \dname{}, where the first one leverages a more centralized approach, while the second one makes use of decentralized resources.

\subsubsection{Best-Antenna}
\begin{figure}
  \centering
  \subfigure[$1^\mathrm{st}$ hop (\cname{}).]{\includegraphics[width=0.32\columnwidth]{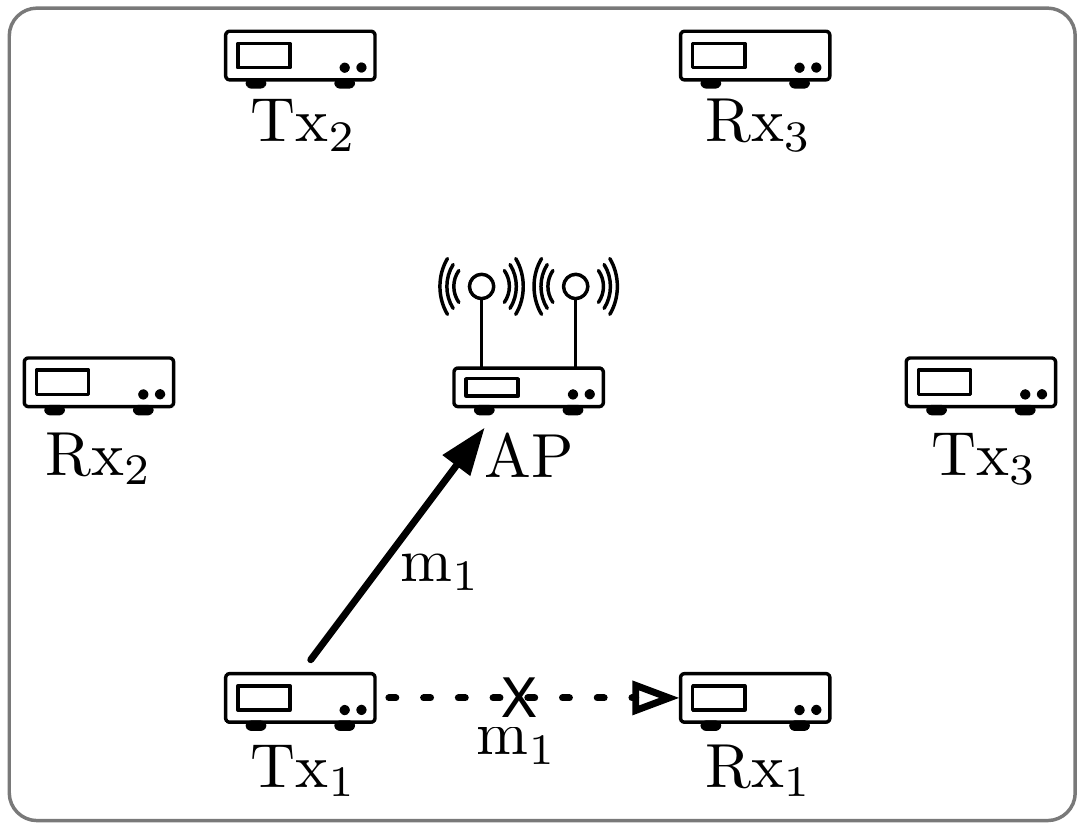}}\quad
  \subfigure[$2^\mathrm{nd}$ hop (\cname{}).]{\includegraphics[width=0.32\columnwidth]{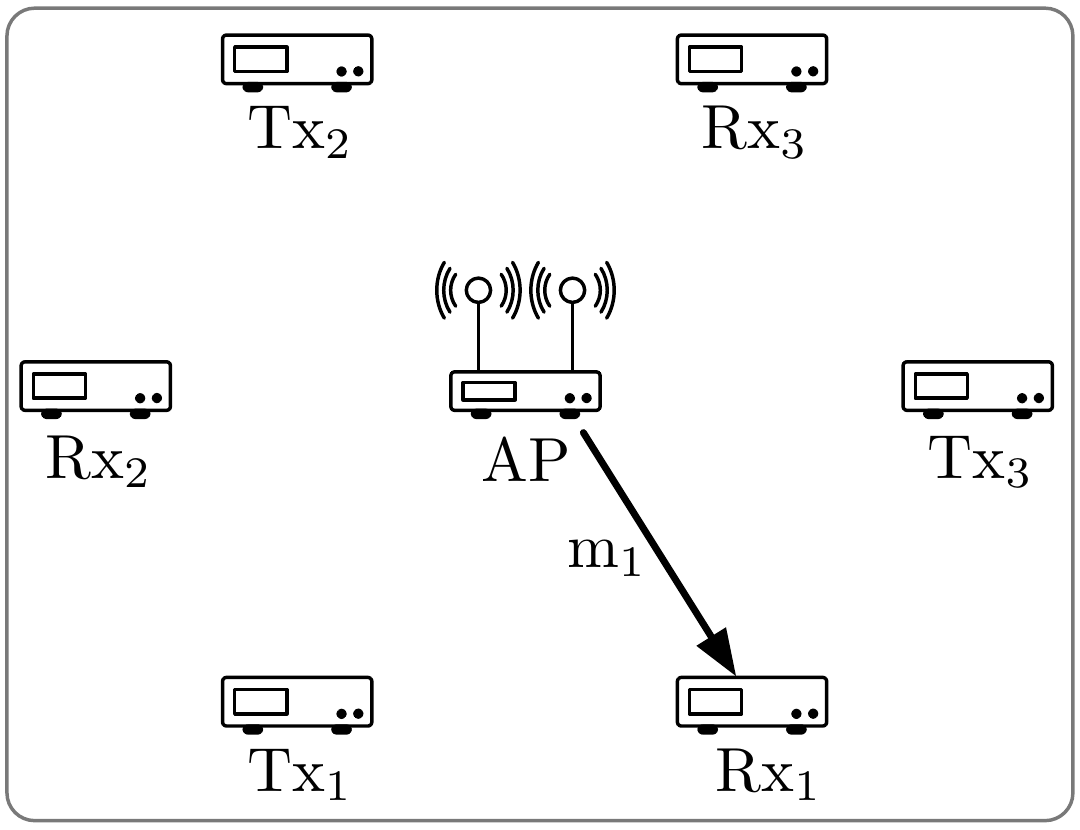}}\quad
  \subfigure[$1^\mathrm{st}$ hop (\dname{}).]{\includegraphics[width=0.32\columnwidth]{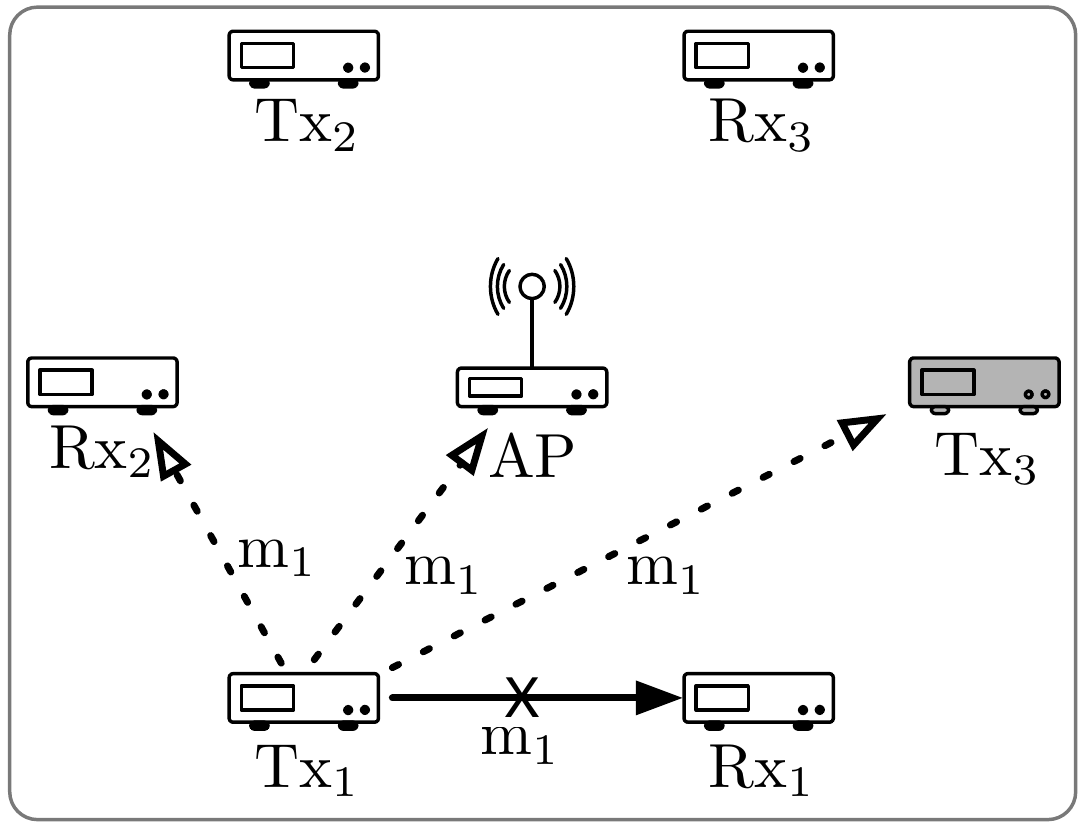}}\quad
  \subfigure[$2^\mathrm{nd}$ hop (\dname{}).]{\includegraphics[width=0.32\columnwidth]{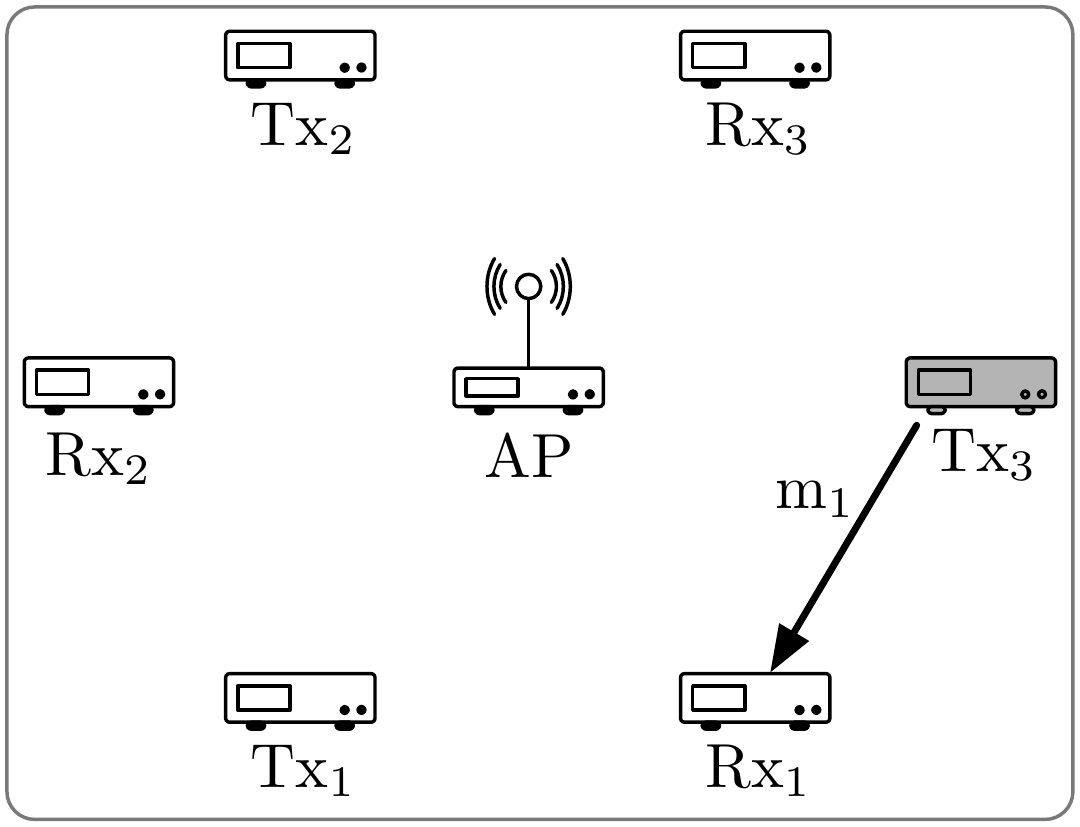}}\quad
  \caption{Example scenario for transmitting a packet m$_1$ from Tx$_1$ to Rx$_1$, illustrating cooperative transmission in \cname{} (a-b) and in \dname{} (c-d). In~(a), the AP schedules an indirect transmission of $m_1$, as the direct link is currently in a bad state, selecting the currently best antenna to receive $m_1$. In~(b), $m_1$ is successfully transmitted from AP to Rx$_1$, again using the currently best antenna for transmission. In (c), three distinct relays overhear $m_1$, while the direct transmission fails. In (d), $m_1$ is relayed by Tx$_3$, which is the selected best relay.}
  \label{fig:relaying}
  \vspace{-5pt}
\end{figure}

This system variant assumes a more asymmetric distribution of hardware resources as it is common in cellular networks, \ie a complex, powerful base terminal and less complex terminals. 
Terminals may thus be limited regarding memory, processing capabilities and transmission antennas in comparison to the \ac{AP} which may have more resources at its disposal, \eg multiple transmission antennas.
Therefore, in this system set-up cooperative transmission is solely performed by the \ac{AP}. 
Transmissions are thus either directly sent from \ac{Tx} to \ac{Rx} or indirectly via the (multi-antenna) \ac{AP}.
The exact decision is performed by a scheduler as discussed in \pref{sec:problem-statement}.
Furthermore, the \ac{AP} uses antenna selection to pick the currently best link for incoming and outgoing transmissions and possibly different antennas on the incoming and outgoing transmission of the same packet.
An example for the relaying in the \cname{} system setup is illustrated in \pref{fig:relaying}~(a-b).

\subsubsection{Best-Relay}
The second system set-up makes full use of the existing distributed resources, assuming that terminals and \ac{AP} have (more or less) comparable hardware characteristics.
Apart from the direct transmission path for a packet between \ac{Tx} and \ac{Rx}, any overhearing terminal in the cell may act as relay to transmit the packet.
More precisely, the \ac{AP} selects for each transmission a direct transmission path or the best available relaying path based on instantaneous \ac{CSI}, \ie by comparing the expected symbol costs for transmitting the packet via these two paths.
Again, the exact scheduling mechanisms is discussed further below.
An example of the system operation in case of the best relay case  \dname{} is illustrated in \pref{fig:relaying}~(c-d).

\subsection{Error Model}
\label{sec:Impact-FB}
A key component impacting any wireless system evaluation is the error model. 
A commonly used outage performance model in wireless systems research is based on the Shannon-Hartley theorem and we refer to this as \acf{IBL} modeling regime.
According to the Shannon-Hartley theorem, the capacity function of a complex channel with \ac{SNR}~$\gamma$, which we denote by ${{\mathcal{C}_\mathrm{IBL}}}\left( \gamma \right)$, is given by ${{\mathcal{C}_\mathrm{IBL}}}({\gamma}) \!=\! {\log _2} ( 1 + \gamma )$ in bits per channel use. 
Following the theorem, a transmission from a sender to a receiver is error-free if ${\mathcal{C}_{{\rm{IBL}}}}\left( \gamma  \right) =\log \left( {1 + \gamma } \right) \ge r  \Leftrightarrow \gamma  \ge 2^{r} - 1$, where $r$ denotes the coding rate (bit/channel use). 
If this requirement is not fulfilled, the packet cannot be decoded correctly, which leads to a packet \emph{outage}.
 The  probability of the outage occurring in an instantaneous  single-hop transmission is given by
\begin{IEEEeqnarray}{RCL}
\label{eq:outage_ibl}
{p}_\mathrm{out} & = & {{\mathbb{P} }}\{ \gamma < 2^{r} - 1\} \quad . 
\end{IEEEeqnarray}
When assuming perfect \ac{CSI} at the sender, \ie the instantaneous $\gamma$ is known, an appropriate rate $r$ can be determined such that $p_\mathrm{out}$ gets zero.
To transmit a packet with size $D$, different values of coding rate $r$ lead to different costs of transmitting symbols, i.e., the  symbol cost (blocklength) results as
\begin{IEEEeqnarray}{RCL}
\label{eq:symbols_cost_IBL}
M=D/r \ge D/{{\mathcal{C}_\mathrm{IBL}}}\left( \gamma \right) \quad . 
\end{IEEEeqnarray}
In other words, under the \ac{IBL} regime a successful transmission of a packet costs a random number of symbols due to the random channel fading.
As a result, when imposing a transmission deadline, the timing/symbol budget might not suffice to reliably convey the packet.
We refer to this error type which is due to the symbol budget limitation as \emph{scheduling error}.

However, as the central goal of our work is to characterize the performance of cooperative systems especially when the target latencies are very short, the Shannon-Hartley theorem becomes a less and less suitable model for the error performance of the links. 
This is due to the fact that it assumes coding blocks of arbitrary length such that the temporarily varying noise averages out. 
While for several thousands of symbols, this assumption might be justified, for low-latency systems it is clearly not the case.
This motivates us to consider a second error model, which we refer to as \acf{FBL} modeling regime. 
In this case, for the real \ac{AWGN} channel,~[8, Theorem 54] derives an accurate approximation of the coding rate for a single-hop transmission system with a finite blocklength.
With a given blocklength $M$, \ac{SNR} $\gamma$, and coding rate $r$, the error probability $\varepsilon$  is given by  
\begin{IEEEeqnarray}{RCL}
\varepsilon & \approx & Q\left( {\frac{{\frac{1}{2}{{\log }_2}\left( {1 + \gamma } \right) - r}}{{\sqrt { {{{V_{{\rm{real}}}}}}/{M}} }}} \right) \quad , 
\end{IEEEeqnarray}
where~$Q (\cdot)$ is the Gaussian Q-function, which is given by $Q\left( w \right) \!=\! {\rm{ }}\int_w^\infty  {\frac{1}{{\sqrt {2\pi } }}} e^{ - t^2 /2} dt$.
In addition, $V_{\text{real}}$ is the  channel dispersion of a real Gaussian channel  given by $V_{\text{real}} = \frac{\gamma }{2}\frac{{\gamma  + 2}}{{{{\left( {1 + \gamma } \right)}^2}}}{\left( {{{\log }_2}e} \right)^2}$.
This result, based on a real \ac{AWGN} channel, has been extended to complex quasi-static fading channel models~\cite{Yang_2014,Gursoy_2013,Peng_2011,Makki_2014,Makki_2015}.
For a single-hop transmission under a quasi-static fading channel and with perfect \ac{CSI} at the sender, the decoding error probability at the receiver is 
\begin{IEEEeqnarray}{RCL}
\label{eq:single_link_error_pro}
\varepsilon & %= &({\gamma},r,m) 
 \approx Q\left( {\frac{{{{\mathcal{C}}_\mathrm{IBL}}({\gamma}) - r}}{{\sqrt {{V_{{\rm{comp}}}}{\rm{/}}M} }}} \right) \quad ,
\end{IEEEeqnarray}
where the  channel dispersion of a complex Gaussian channel is twice the one of a real Gaussian channel, i.e.,
${V_{{\rm{comp}}}} \!=\! 2{V_{{\rm{real}}}} %\!=\! \gamma \frac{{\gamma  \!+\! 2}}{{{{\left( {1 \!+\! \gamma } \right)}^2}}}{\left( {{{\log }_2}e} \right)^2}
 \!=\! ( {1 \!-\! \frac{1}{{{{\left( {1 + \gamma } \right)}^2}}}} ){\left( {{\log_2}e} \right)^2} $.
These approximations have been shown to be tight for sufficiently large values of $M$~\cite{Verdu_2010, Polyanskiy_2011, Yang_2014}.
In the remainder of the paper, we consider sufficiently large values of $M$ for each transmission.

Comparing \pref{eq:outage_ibl} with \pref{eq:single_link_error_pro}, the difference between the two error models becomes evident: 
Errors under the \ac{IBL} regime are solely caused by scheduling, while the error probability under the \ac{FBL} regime is influenced by both the scheduling and the decoding due to finite blocklengths.

\subsection{Overhead of Acquisition of  \ac{CSI}}
\label{sec:csi-acquisition}

In both the \cname{} and the \dname{} system variant, the \ac{AP} uses perfect \ac{CSI} to schedule the transmissions.
In practice, this implies that for each considered link, the current link conditions must first be determined  and then communicated to the \ac{AP}.
The former manifests  as \emph{time} overhead, which in practical systems corresponds to a reference signal preceding the packet transmission.
The latter manifests as \emph{communication} overhead, as the link information must be transmitted to the \ac{AP}.
A possible approach is to piggyback this information at the end of payload packets in regular transmissions, which are overheard by the \ac{AP}.
All links from and to the \ac{AP} can be directly estimated by the \ac{AP}, leading to no communication overhead for these links.
For a single link, we define $\alpha$ as the duration of the reference signal in symbols, while $\beta$ indicates the number of bits required to represent the link quality and thus corresponds to the communication overhead per link.
The total number of symbols to estimate the qualities of all links depends, in both system variants, on the number of transmissions per frame $N$, leading to $N \cdot \alpha$.
The total communication overhead, however, depends on the number of considered links and therefore differs for each system variant.

In \cname{}, packets are either transmitted directly between \ac{Tx} and \ac{Rx} or indirectly via the \ac{AP}.
All relay links can thus be estimated by the \ac{AP} and therefore do not increase the communication overhead.
For the direct transmissions, the respective links are estimated by the terminals and consequently this information must be conveyed to the \ac{AP}.
Thus, a total of $N$ links must be characterized, leading to a total communication overhead of $N \cdot \beta$.
Hence, the size of a single packet increases to $\unit[D + \beta]{bits}$.

In \dname{}, any terminal including the \ac{AP} may potentially act as relay, leading to a fully connected network.
However, as links from and to the \ac{AP} can be excluded, the total number of considered links is $\frac{n(n-1)}{2}$.
Assuming a fixed order in which the link qualities are reported to the \ac{AP}, the total message overhead for the decentralized system variant is $\frac{n(n-1)}{2} \cdot \beta$.
This leads to a packet size of $\unit[D + \frac{(n-1)}{2} \cdot \beta]{bits}$.

\subsection{Scheduling and Problem Statement}
\label{sec:problem-statement}
The main objective of this work is to study how the packet error rate (PER) behaves for a multi-terminal wireless transmission system incorporating cooperation with a stringent time deadline, \ie in each transmission cycle there is only a finite number of transmission symbols $S$ that must be shared by the associated terminals.
We consider two fundamental design options regarding the relaying process to study the system performance when using centralized resources for relaying compared to the use of decentralized resources.
Under both the \ac{IBL} and the \ac{FBL} regime, to reduce the error probability the \ac{AP} leverages cooperative relaying in combination with perfect \ac{CSI} to select reliable transmission paths, minimizing for each transmission the number of needed symbols.
The difference is that for calculating the cost of symbols under the \ac{IBL} modeling regime we base the derivations on \pref{eq:symbols_cost_IBL}, while for the \ac{FBL} modeling regime it is according to \pref{eq:single_link_error_pro}.

For a terminal~$i$, under the \ac{IBL} and the \ac{FBL} regime, the symbol cost of a direct transmission is denoted by $M_{\mathrm{D},i}$ and the cost of relaying is denoted by $M_{\mathrm{R},i}$.
The \ac{AP} selects the option with the minimal costs, \ie $M_{\mathrm{min},i}= \min\{M_{\mathrm{R},i}, M_{\mathrm{D},i}\}$.
Note that a relay path consists of two hops, the link from \ac{Tx} to relay, denoted by $\mathrm{R1}$, and the link from relay to \ac{Rx}, denoted by $\mathrm{R2}$, so that $M_{\mathrm{R},i} =M_{\mathrm{R1},i}+M_{\mathrm{R2},i}$.
In both regimes, it is possible that due to random fading the number of symbols $S$ does not suffice to reliably convey all $N$ packets.
In this case, the first packets are scheduled until $S$ is exceeded and the remaining packets are dropped.
The probability that only the first $i$ packets are scheduled is denoted by $p_i$.
Hence, the probability of packet $i$ not being scheduled is $1-{p_i}$.

So far, we have introduced the scheduling model for the system. 
In the following, we give details on the \ac{PER} performance under the \ac{IBL} and the \ac{FBL} regime, respectively. 
The \ac{PER} under the \ac{IBL} regime is fully subject to the probability of scheduling errors, \ie $1-{p_i}, i=1,\ldots,N$. 
In particular, the average \ac{PER} over $N$ packets in the \ac{IBL} regime is
\begin{IEEEeqnarray}{RCL}
\label{eq:FinalShannon_PER}
 {\rm{PER}}_{{\rm{IBL}}}  = \frac{1}{N}\sum\limits_{i = 1}^N {\left\{ 1 - {p_i}   \right\}} \quad .
\end{IEEEeqnarray}

Under the \ac{FBL} regime, in addition to scheduling errors, decoding errors also occur at the receiver due to limited blocklengths.
Thus, the \ac{AP} considers a certain target decoding error probability $\varepsilon^*$ when allocating the symbols of a packet in a single-hop transmission.
This target error probability influences the overall reliability of a transmission.
With probability $ {{\mathbb{P} }}\left\{ M_{\mathrm{R},i} \ge M_{\mathrm{D},i} \right\} $, the target error probability is $\varepsilon^*$. 
In turn, when relaying a packet from transmitter terminal~$i$ with $ {{\mathbb{P} }} \left\{ {{M_{{\rm{R}},i}} < {M_{{\rm{D}},i}}} \right\}$, the target error probability of each link yields a two-hop target error probability of $1-(1-\varepsilon^*)^2  =  2{\varepsilon ^*}-  {(\varepsilon ^*)^2} \approx  2{\varepsilon ^*}$\footnote{Considering reliable wireless systems with $\varepsilon^* \! \ll \! 10^{-1}$, thus, $2{\varepsilon ^*} \! \gg \! {(\varepsilon ^*)^2}$.}.
Thus, the expected error probability for a scheduled packet $i$ is  $\varepsilon _{\mathrm{ave},i}^*=  {{\mathbb{P} }}\left\{ M_{\mathrm{R},i} \ge M_{\mathrm{D},i} \right\} \cdot \varepsilon ^* +  {{\mathbb{P} }} \left\{ M_{\mathrm{R},i} < M_{\mathrm{D},i} \right\} \cdot 2 \varepsilon^*$.
The combined \ac{PER} of a packet $i$ under the \ac{FBL} regime is then given by
\begin{IEEEeqnarray}{RCL}
\label{eq:PER_j}
{\rm{PER}}_{{\rm{FBL}},i} = 1 - p_i + p_i \cdot \varepsilon _{\mathrm{ave},i}^* \quad .
\end{IEEEeqnarray}
Finally, under the \ac{FBL} regime the \ac{PER} over all $N$ packets results to
\begin{IEEEeqnarray}{RCL}
\label{eq:Final_PER}
 {\rm{PER}}_{{\rm{FBL}}} = \frac{1}{N}\sum\limits_{i = 1}^N {\rm{PER}}_{{\rm{FBL}},i}  
= \frac{1}{N}\sum\limits_{i = 1}^N {\left\{ {1 - {p_i} + {p_i}\varepsilon _{{\rm ave} ,i}^*} \right\}} \quad .
\end{IEEEeqnarray}

By comparing~the above \ac{PER} models of the \ac{IBL} and the \ac{FBL} regime, the one under the \ac{IBL} regime can be seen as a special case of the one under the \ac{FBL} regime, where $m \to  + \infty$ and ${\varepsilon ^*} \to 0$. 
In particular, \pref{eq:FinalShannon_PER} can be obtained by substituting ${\varepsilon ^*} =0$ into \pref{eq:Final_PER}.

Given this general model for the \ac{PER} performance, the following questions are addressed in the further course of this paper:
(i)~What is the exact analytical performance model of the proposed systems especially under the \ac{FBL} regime?
(ii)~What are the performance properties of the considered system variants, \ie how do they scale with respect to the overhead, the load, the resource budget, and the target error probability?
(iii)~How is this scaling behavior different when analyzing the two systems under the \ac{IBL} or the \ac{FBL} modeling regime?

%% file: 05_error_probability.tex
\section{Packet Error Probability in the Finite Blocklength Regime}
\label{sec:finite}
The receiver \acp{SNR} are random variables subject to channel fading.
The cost of reliably transmitting a packet from a terminal~$i$ to a terminal~$k$, in terms of symbols, thus varies over time.
We characterize this random cost by the \ac{PDF} ${f_{{M_{i,k}}}} \left( m  \right)$. % for short..
Consequently, the \acp{PDF} of $M_{{\rm min},i}$, $M_{{\rm D},i}$, and $M_{{\rm R},i}$ (\cf \pref{sec:problem-statement}) can be given by ${{f_{{M_{{\rm min},i}}}}(m)}$, ${{f_{{M_{{\rm D},i}}}}(m)}$, and ${{f_{{M_{{\rm R},i}}}}(m)}$, respectively. % In particular, since that 
In the following, we first focus on ${{f_{{M_{{\rm min},i}}}}(m)}$ and on the average \ac{PER} of the considered system for given ${{f_{{M_{{\rm R},i}}}}(m )}$ and ${{f_{{M_{{\rm D},i}}}}(m)}$, $i=0,\ldots,N$. 
Afterward, we derive the \acp{CDF} ${{F_{{M_{{\rm R},i}}}}(m )}$ and ${{F_{{M_{{\rm D},i}}}}(m)}$ for both relaying strategies.

\subsection{Average PER}
The \acp{CDF} of $M_{{\rm D},i}$ and $M_{{\rm R},i}$ are given by ${{F_{{M_{{\rm R},i}}}}(m )}$ and ${{F_{{M_{{\rm D},i}}}}}(m)$, respectively, then the \ac{CDF} of $M_{{\rm min},i}$ can be derived as follows
\begin{IEEEeqnarray}{RCL}
\label{eq:min_relay_vs_direct}
{F_{{M_{{\rm min},i}}}}(m) & = & 1 - \left( {1 - {F_{{M_{{\rm R},i}}}}(m) } \right)\left( {1 - {F_{M_{{\rm D},i}}}(m)} \right) \quad .
\end{IEEEeqnarray}
Hence, the \ac{PDF} of $M_{{\rm min},i}$ is given by
\begin{IEEEeqnarray}{RCL}
\label{eq:m-min-pdf}
{f_{{M_{{\rm{min}},i}}}}(m) & = & {F_{{M_{\text R},i}}}(m){f_{{M_{{\text D},i}}}}(m)
 + \left( {1 - {F_{{M_{{\text D},i}}}}(m)} \right){f_{{M_{{\text R},i}}}}(m) \quad .
\end{IEEEeqnarray}

Recall that a total of $N$ packets need to be transmitted during a frame while the minimal cost for transmitting a packet from terminal~$i$ is $M_{{{\min}},i}, i \!=\! 1,\ldots,N$, which are i.i.d.
Then, the \ac{PDF} of the sum of the  cost of transmitting  all $N$ packets  $M_{\rm sum} =\sum\limits_{i = 1}^N {{M_{{\rm{min}},i}}} $ is given based on \pref{eq:m-min-pdf} as
\begin{IEEEeqnarray}{RCL}
f_{M_{\rm sum}} (m) =   {{f_{{M_{{{ {\min}},1}}}}}(m) \otimes  \ldots  \otimes {f_{{M_{{{\min},N}}}}}(m)} \quad , 
\end{IEEEeqnarray}
where $\otimes$ is the convolution function.

The probability that the first $n$ packets are successfully transmitted in a frame with total blocklength $S$ is given by 
\begin{IEEEeqnarray}{RCL}
\label{eq:error_of_scheduling_each_j}
p_k & = &F_{M_{\rm sum}} (S) \quad .
\end{IEEEeqnarray}
To derive the average \ac{PER} over all $N$ packets, denoted by ${\rm{PER}_\mathrm{FBL}}$, the target error probability $\varepsilon^*$ needs to be considered. 
For a scheduled packet at terminal~$i$ with a probability of $ {{\mathbb{P} }} \left\{ {{M_{{\rm{R}},i}} \ge {M_{{\rm{D}},i}}} \right\} = \sum_{m = 1}^{+ \infty} {{F_{{M_{{\rm{D}},i}}}}(m){f_{{M_{{\rm{R}},i}}}}(m)} $ the transmission error probability is $\varepsilon^*$, while with a probability of $ \mathbb{P} \left\{ M_{{\rm{R}},i} < M_{{\rm{D}},i} \right\} = \sum_{m = 1}^{ + \infty } {{F_{{M_{{{\rm{R}},i}}}}}(m){f_{{M_{{{\rm{D}},i}}}}}(m)}$ the transmission error probability is $2\varepsilon^*$. 
Hence, the expected error probability for a scheduled packet $i$ is given by
\begin{IEEEeqnarray}  {RCL}
\label{eq:error_of_scheduled_packet}
\varepsilon _{{\rm ave},i }^* =  \sum\limits_{m = 1}^{ + \infty } {{F_{{M_{{\rm{D}},i}}}}(m){f_{{M_{{\rm{R}},i}}}}(m)} \ {\varepsilon ^*}  
 + \sum\limits_{m = 1}^{ + \infty } {{F_{{M_{{\rm{R}},i}}}}(m){f_{{M_{{\rm{D}},i}}}}(m)} \ 2{\varepsilon ^*} \quad .
\end{IEEEeqnarray}
Then, the combined \ac{PER} for the $i$th packet and the average \ac{PER} over all $N$ packets can be obtained by \pref{eq:PER_j} and \pref{eq:Final_PER}.

So far, we derived the \ac{PER} under the \ac{FBL} regime with given \acp{PDF} $M_{{\rm R},i}$ and $M_{{\rm D},i}$.
In the following, we focus on the derivation of these \acp{PDF} considering direct transmissions, \dname{}, and \cname{}.

\subsection{Distribution of the Transmission Blocklengths}
\label{sec:distributions-blocklengths}

According to \pref{eq:single_link_error_pro}, the error probability of a single-hop transmission with packet size $D$ and blocklength $M$ is 
\begin{IEEEeqnarray}{RCL}
\label{eq:Polyanskiy_errorpro}
\varepsilon = 
Q\left( {\frac{{{\mathcal{C}_\mathrm{IBL}}(\gamma ) - D/M}}{{{{\log }_2}e \ \sqrt {\left( {1 -   {{{{\left( {1 + \gamma } \right)}^{-2}}}}} \right){\rm{/}}M} }}} \right) \quad .
\end{IEEEeqnarray}
If the error probability of each transmission is required to be lower than ${{\varepsilon ^*}}<0.5$, then the minimal blocklength $M^*$ satisfies
\begin{IEEEeqnarray}{RCL}
\label{eq:Polyanskiy_error}
{\varepsilon ^*} = Q\left( {\frac{{ \mathcal{C}_\mathrm{IBL}(\gamma ) - D/{M^*}}}{{{{\log }_2}e \ \sqrt {\left( {1 - \frac{1}{{{{\left( {1 + \gamma } \right)}^2}}}} \right){\rm{/}}{M^*}} }}} \right) \quad .
\end{IEEEeqnarray}
In particular, we further have
\begin{IEEEeqnarray}{RCL}
 {\left( {\sqrt {{M^*}} } \right)^2} -  v\sqrt {{M^*}}  - D/\mathcal{C}_\mathrm{IBL}(\gamma ) = 0 \quad ,
\end{IEEEeqnarray}
where $v  = {Q^{ - 1}}\left( {{\varepsilon ^*}} \right) \frac{{{{\log }_2}e\sqrt {\left( {1 - \frac{1}{{{{\left( {1 + \gamma } \right)}^2}}}} \right)} }}{{\mathcal{C}_\mathrm{IBL}(\gamma )}}$, 
which leads to
\begin{IEEEeqnarray}{RCL}
\sqrt {{M^*}}  = \sqrt {\frac{D}{ {\mathcal{C}_\mathrm{IBL}(\gamma )} } + {{\left( {\frac{{ v}}{{{{2  }} }}} \right)}^2}}  + \frac{v}{{{{2 }}}} \quad .
\end{IEEEeqnarray}
Finally, this results in a minimal blocklength $M^*$ of
\begin{IEEEeqnarray}{RCL}
{M^*} = \frac{D}{ {\mathcal{C}_\mathrm{IBL}(\gamma )} } + \frac{1}{2}{v^2} + v\sqrt {\frac{D}{ {\mathcal{C}_\mathrm{IBL}(\gamma )} } + {{\left( {\frac{v}{{\rm{2}}}} \right)}^2}} \quad .
\end{IEEEeqnarray}

Obviously, $M^*$ is a function of $\gamma$ and $v$, while $v$ is a function of $\gamma$.
Consequently, $M^*$ is a function of $\gamma$. 
We denote this function as $g(\cdot)$, \ie $M^* = g (\gamma)$. 
Then, the corresponding inverse function is given by $\gamma= g^{-1} (M^* )$. 
Based on the channel gain distribution in \pref{eq:gamma_distribution}, the \ac{CDF} of $M^*$ is
\begin{IEEEeqnarray}{RCL}
{F_{{M^*}}}\left(m, \overline{\gamma} \right) = \int\limits_{z  \in \Omega } {{p}\left( z  \right)dz }  = \int\limits_{0}^{{g^{ - 1}}\left( m \right)/ \overline{\gamma}} {{p }\left( z  \right)dz } \quad ,
\end{IEEEeqnarray}
where $\Omega  = \left\{ {z :{M^*}\left( z \overline{\gamma} \right) \le m} \right\}$.
Then the \ac{PDF} of $M^*$ of a single-hop link with average channel gain $\overline{\gamma}$ is
\begin{IEEEeqnarray}{RCL}
\label{eq:PDF_single_hop}
{f_{{M^*}}}\left( {m,\overline{\gamma}} \right) = \frac{{\partial {F_{{M^*}}}\left( m \right)}}{{\partial m}} = \frac{{{p_{\overline{\gamma} }}\left( {{g^{ - 1}}\left( m \right)} \right)}}{{\frac{{\partial g\left( m \right)}}{{\partial m}}}} \quad .
\end{IEEEeqnarray}
Based on \pref{eq:PDF_single_hop}, the \ac{PDF} of the cost of transmitting a packet via the direct link between terminal~$i$ and~$k$ can be expressed as ${f_{{M^*}}}\left( {m,\overline{\gamma}_{i,k} } \right) $.
 
When applying the best relay strategy, where the \ac{AP} selects the terminal with the lowest transmission cost to act as relay, the \ac{PDF} of the lowest cost is given by%\fmms{we need to fix the indices $i$, $j$, and $k$.} 
\newtheorem{theorem}{Lemma}
\begin{theorem}
	\label{le:relay-cost}
{	 Under the best relay strategy, the \ac{PDF} of the minimal cost of transmitting packet $i$ via the best relay over $J$ relay candidates is given by
\begin{IEEEeqnarray}{RCL}
\label{eq:PDF_Best_R_FB}
{f_{{M_{{\rm{R}},i}}}}(m) = \sum\limits_{j = 1}^J {\prod\limits_{\substack{s = 1\\ s \ne j}}^J {{f_{M_{{\rm{R}},j}^{i - k}}}(m)\left( {1 - {F_{M_{{\rm{R}},s}^{i - k}}}(m)} \right)} } \quad .
\end{IEEEeqnarray}}
\end{theorem}

\begin{IEEEproof} 
Under the best relay strategy, if  terminal~$j$ acts as a relay, the \acp{PDF} of $m_{R1,i}$ and $m_{R2,i}$ are ${f_{{M^*}}}\left( {m,\overline{\gamma}_{i,j} } \right) $ and ${f_{{M^*}}}\left( {m,\overline{\gamma}_{j,k} } \right) $. 
Hence, the \ac{PDF} of the sum of the cost of the two hops is given by
\begin{IEEEeqnarray}{RCL}
\label{eq:PDF_each_R_FB}
{f_{{M_{{\rm R},j}^{i-k}}}}(m)  ={f_{{M^*}}}\left( m, \overline \gamma_{i,j}  \right) \otimes {f_{{M^*}}}\left( m, \overline \gamma_{j,k}  \right) \quad ,
\end{IEEEeqnarray}
with \ac{CDF} ${F_{M_{{\rm{R},j}}^{i-k}}}(m) = \int_0^m {f_{M_{{\rm{R}},j}^{i-k}}}(t) \, dt$.
Note that in \dname{} only the terminal with the smallest costs is selected to relay the packet.
The \ac{CDF} of the minimal cost of transmitting packet $i$ via one of the $J$ relay candidates is given by
\begin{IEEEeqnarray}{RCL}
{F_{{M_{{\rm{R}},i}}}}(m) = 1 - \prod\limits_{j = 1}^J {\left( {1 - {F_{{M_{{\rm{R}},j}^{i-k}}}}(m)} \right)} \quad .
\end{IEEEeqnarray}
Finally, we have the \ac{PDF} of the minimal blocklength as shown in \pref{le:relay-cost}.
\end{IEEEproof} 
Hence, the \ac{PER} of the best relay strategy can be obtained by substituting \pref{le:relay-cost} into \pref{eq:Final_PER} and \pref{eq:min_relay_vs_direct}.
On the other hand, when applying \cname{} only the \ac{AP} may act as relay.
Therefore, the \ac{PDF} of the cost of the first and the second hop of the transmission from terminal~$i$ to terminal~$k$ via an antenna of the \ac{AP} is given by ${f_{{M^*}}}\left( {m,\overline{\gamma}_{i,0} } \right) $ and ${f_{{M^*}}}\left( {m,\overline{\gamma}_{0,k} } \right) $.
\begin{theorem}
\label{le:antenna-cost}
Under the best antenna strategy, the \ac{PDF} of the minimal cost of transmitting the packet for terminal $i$ via one of the~$J$ antennas of the \ac{AP} is given by
\begin{IEEEeqnarray}{RCL}
\label{eq:PDF_Best_A_FB}
{f_{{M_{{\rm R},i}}}}(m) = f_{{M_{{\rm R}1,i}}}(m) \otimes f_{{M_{{\rm R}2,i}}}(m) \quad ,  
\end{IEEEeqnarray}
where
\begin{IEEEeqnarray}{RCL}
\label{eq:PDF_each_A_FB}
\begin{aligned}
f_{{M_{{\rm R}1,i}}}(m) = &  J \left(  1 - {F_{M^*}(m,\bar \gamma _{i,0}  )} \right)^{J - 1}f_{M^*}(m,\bar \gamma _{i,0}) \quad , \\ 
f_{{M_{{\rm R}2,i}}}(m) = & J \left(  1 - {F_{M^*}(m,\bar \gamma _{0,k}  )} \right)^{J - 1}f_{M^*}(m,\bar \gamma _{0,k}  ) \quad .
\end{aligned}
\end{IEEEeqnarray}
\end{theorem}
\begin{IEEEproof}
Recall that the best antenna out of $J$ antennas for the first hop and the best one out of $J$ antennas for the second hop are selected.
$f_{{M_{{\rm R}1,i}}}(m) $ and $f_{{M_{{\rm R}2,i}}}(m) $ in \pref{eq:PDF_each_A_FB} are actually the \ac{PDF} of the minimal costs for the first and the second hop via the \ac{AP}. 
Then, ${f_{{M_{{\rm R},i}}}}(m) $ is the \ac{PDF} of the sum of ${{M_{{\rm R}1,i}}}(m) $ and ${{M_{{\rm R}2,i}}}(m) $, as given in \pref{eq:PDF_Best_A_FB}. 
According to \pref{eq:Final_PER} and \pref{eq:min_relay_vs_direct}, the corresponding \ac{PER} under the best antenna strategy can be obtained.
\end{IEEEproof} 
Until now, the \acp{PER} of the two system variants have been studied. 
Under these two variants, packets are either transmitted directly or via a relay.
The key difference is that in \dname{} one terminal is selected as relay, while in \cname{} the multi-antenna \ac{AP} acts as relay.
For both variants, we state the following theorem.
\newtheorem{theorem-non}{Theorem}
\begin{theorem-non}
	\label{th:convex-error}
	For the \ac{FBL} modeling regime and for the two considered systems, the average \ac{PER} of a single packet~$i$, denoted by ${\rm{PER}}_{{\rm{FBL}},i}$ with $i=1,\ldots,N$, as well as the average system \ac{PER} over all $N$ packets transmitted per frame, denoted by ${\rm{PER}}_{{\rm{FBL}}}$, are both convex in the target decoding error probability $\varepsilon^*$. 
\end{theorem-non}
\begin{IEEEproof} {See Appendix A.}
\end{IEEEproof}

%% file: 06_outage_IBL_regime.tex
\section{Packet Error Probability in the Infinite Blocklength Regime}
\label{sec:infinite}

Recall that under the \ac{IBL} regime, a single-hop transmission is error free if $ \mathcal{C}_\mathrm{IBL}\left( \gamma  \right) \! = \! \log \left( {1 + \gamma } \right) \! \ge \! {\frac{D}{M}}   \Leftrightarrow \gamma \! \ge \! {2^{\frac{D}{M}}} \! - \! 1$.
Hence, the minimal blocklength cost $M^*$ for successfully transmitting a packet is the realization of a random variable.  
Considering that it is required to transmit $N$ packets per frame within a fixed frame length of $S$ symbols, the transmission error of the considered system in the \ac{IBL} regime is fully subject to scheduling, \ie the sum of the minimal costs for transmitting $N$ packets may be larger than $S$. 
Since we assume a block-fading Rayleigh channel, the \ac{CDF} of the minimal blocklength $M^*$ for transmitting a packet of size~$D$ via a single-hop transmission with average \ac{SNR} $\overline{\gamma}$ is given by
\begin{IEEEeqnarray}{RCL}
\label{eq:CDF_single_hop_cost_IFB}
{F_{M^*}}\left( {m,\overline{\gamma} } \right) & = & \Pr \{ M^* \le m\}    
 =  \Pr \{ \gamma  \ge {2^{\frac{D}{m}}} - 1\} 
  =   \exp \left[ { - \frac{1}{ \overline{\gamma}}} \left( {{2^{\frac{D}{m}}} - 1} \right) \right]\quad .
\end{IEEEeqnarray} 
The \ac{PDF} of the minimal cost of a single-hop transmission with average \ac{SNR} $\overline{\gamma}$ is then
\begin{IEEEeqnarray}{RCL}
\label{eq:PDF_single_hop_cost_IFB}
{f_{M^*}} \! \left( {m,\overline{\gamma}} \right){\rm{ }} = {\rm{ }}\exp \! \left[ { { - \frac{1}{ \overline{\gamma}}}\left( {{2^{\frac{D}{m}}} - 1} \right)} \right] \! \cdot \! \frac{{{2^{\frac{D}{m}}}}}{{\overline{\gamma} }} \! \cdot \! \frac{{D\ln 2}}{{{m^2}}} \quad .
\end{IEEEeqnarray}
Then,  the average \ac{PER} over all $N$ packets can be obtained by \pref{eq:FinalShannon_PER}.
Note that the \ac{IBL} regime can be seen as a special case of the \ac{FBL} regime, where $m \to  + \infty$ and ${\varepsilon ^*} \to 0$.
Hence, the derivations in the previous section still hold in the \ac{IBL} regime.
In particular, we can derive $p_i$ for \dname{} by substituting \pref{eq:CDF_single_hop_cost_IFB} and \pref{eq:PDF_single_hop_cost_IFB} into \pref{eq:Final_PER}, \pref{eq:min_relay_vs_direct}, \pref{eq:PDF_Best_R_FB}, and \pref{eq:PDF_each_R_FB}.
Similarly, for \cname{}, the \ac{PER} can be obtained by substituting \pref{eq:CDF_single_hop_cost_IFB} and \pref{eq:PDF_single_hop_cost_IFB} into \pref{eq:Final_PER}, \pref{eq:min_relay_vs_direct}, \pref{eq:PDF_Best_A_FB}, and \pref{eq:PDF_each_A_FB}.

%% file: 07_evaluation.tex
\section{Performance Evaluation}
\label{sec:eval}
In this section, we first empirically validate the correctness of our theoretical model by simulations.
In this regard, we are especially interested in validating \pref{th:convex-error} (\cf \pref{sec:distributions-blocklengths}) to discuss the role of the selected target error probability on the \ac{PER}.
Subsequently, we numerically evaluate the system performance with the proposed models for the \ac{PER}.
Our aim is to analyze under which conditions ultra-high reliability (\ac{PER}$<\!\!10^{-9}$) with ultra-low latencies (below $\unit[1]{ms}$) can be achieved through cooperative transmission and how the proposed systems differ in their performance when considering the \ac{IBL} or \ac{FBL} modeling regime.
For different setups, we thus compare the results under the \ac{FBL} and the \ac{IBL} regime to illustrate the impact of finite blocklengths, which is typically not considered in related work, and finally also consider the scaling behavior.
For both the validation and the evaluation part, we consider the parameterization of the system model shown in \pref{tab:params}.
\begin{table}
	\centering
	\caption{Validation/Evaluation Parameters.}
	\label{tab:params}
	\footnotesize
	\begin{tabular}{@{}cll@{}}
		\toprule
		Symb. & Value & Description \\
		\midrule
		& & \\[-3mm]
		$B$ & \unit[5]{MHz} & Channel bandwidth.\\
        $S$ & 5000 & Total amount of symbols per frame.\\
        $N$ & 5 & Number of transmissions per frame.\\
        $\alpha$ & $\nicefrac{S}{100}$ & Required symbols to estimate the link quality.\\
        $\beta$ & \unit[8]{bits} & Required bits to represent the link quality.\\
		$D$ & $\unit[128]{bit} + N \beta$  & Packet size in \direct{} / \cname{}.\\
		$D$ & $\unit[128]{bit} + \frac{N-1}{2} \beta$  & Packet size in \dname{}.\\
		$\overline{\gamma}$ & \unit[15]{dB} & Average \ac{SNR} at the receiver.\\
		\bottomrule
	\end{tabular}
\end{table}

\subsection{Simulative Validation}
\label{sec:validation}

We empirically validate $\mathrm{PER}_\mathrm{FBL}$ (\cf \pref{eq:Final_PER}) for \direct{}, \dname{}, and \cname{} by simulations.
Therefore, we generate random instances of the receiver \ac{SNR}, which is exponentially distributed around the average.
The channel instances are used to calculate, for each transmission, the minimal blocklength $M^*$ according to the considered model and subsequently to compute the respective \ac{PER}.
The simulation is implemented in \texttt{Python} using \texttt{NumPy}.
For each data point, we generate at least $10^8$ transmission frames to be able to empirically observe the expected \ac{PER}.
Note that in the case of \dname{} and \cname{}, we set the number of available relays/antennas to one and two, leading to \acp{PER} that can be verified by simulations in a reasonable amount of time.

\begin{figure}
\centering
\includegraphics[width=0.6\columnwidth]{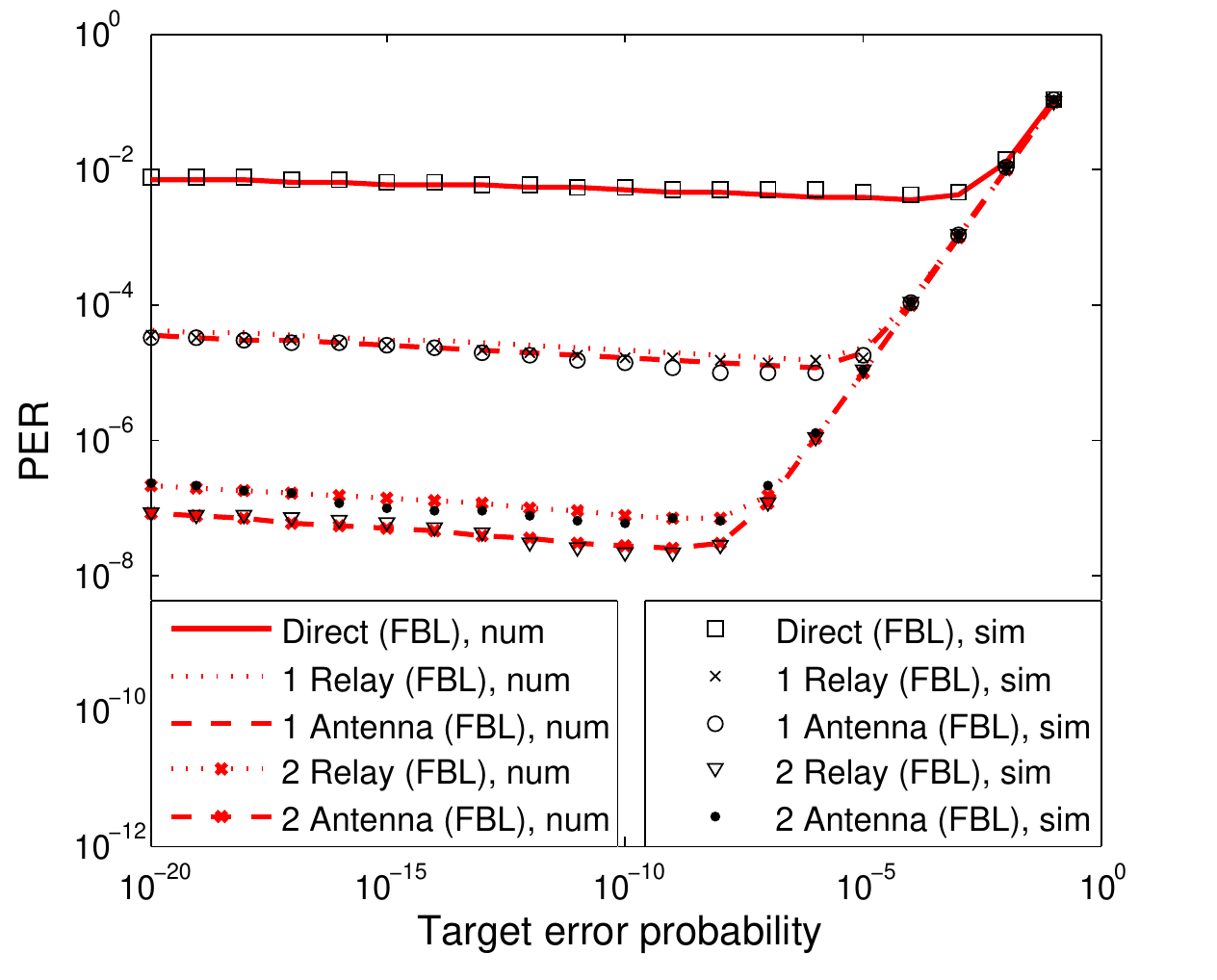}
\caption{Simulative validation of \direct{}, \dname{}, and \cname{} under the \ac{FBL} regime varying the target error probability $\epsilon^*$.}
\label{fig:validation}
\end{figure}
The corresponding results are illustrated in \pref{fig:validation}.
Markers indicate simulation results, while lines indicate the respective numerical results for comparison.
We see that the simulation accurately matches the numerical results as only small deviations are observed due to a finite number of samples in the simulation.
Moreover, these results confirm \pref{th:convex-error} (\cf \pref{sec:distributions-blocklengths}), showing that the $\mathrm{PER}_\mathrm{FBL}$ is convex in~$\epsilon^*$.
In general, introducing a higher cooperative diversity, \ie with more antennas/relays, leads to a lower $\mathrm{PER}_\mathrm{FBL}$ at the optimum.
Once the optimum is reached, $\mathrm{PER}_\mathrm{FBL}$ increases moderately with a lower~$\epsilon^*$ for the considered parametrization.
This actually already reveals a key trade-off in the considered systems between the scheduling error and the decoding error floor. 
The plot strongly motivates to rather choose the decoding error conservatively, leading to a higher impact due to the scheduling error in comparison to the optimal point of operation. 
We provide more details on this below.

\subsection{Finite Versus Infinite Blocklength Regime} 

\renewcommand{\mycap}{Varying the packet size $D$ for \cname{} and \dname{}.}
\doublefig{fig:2regime_Packet}
{\cname{}} {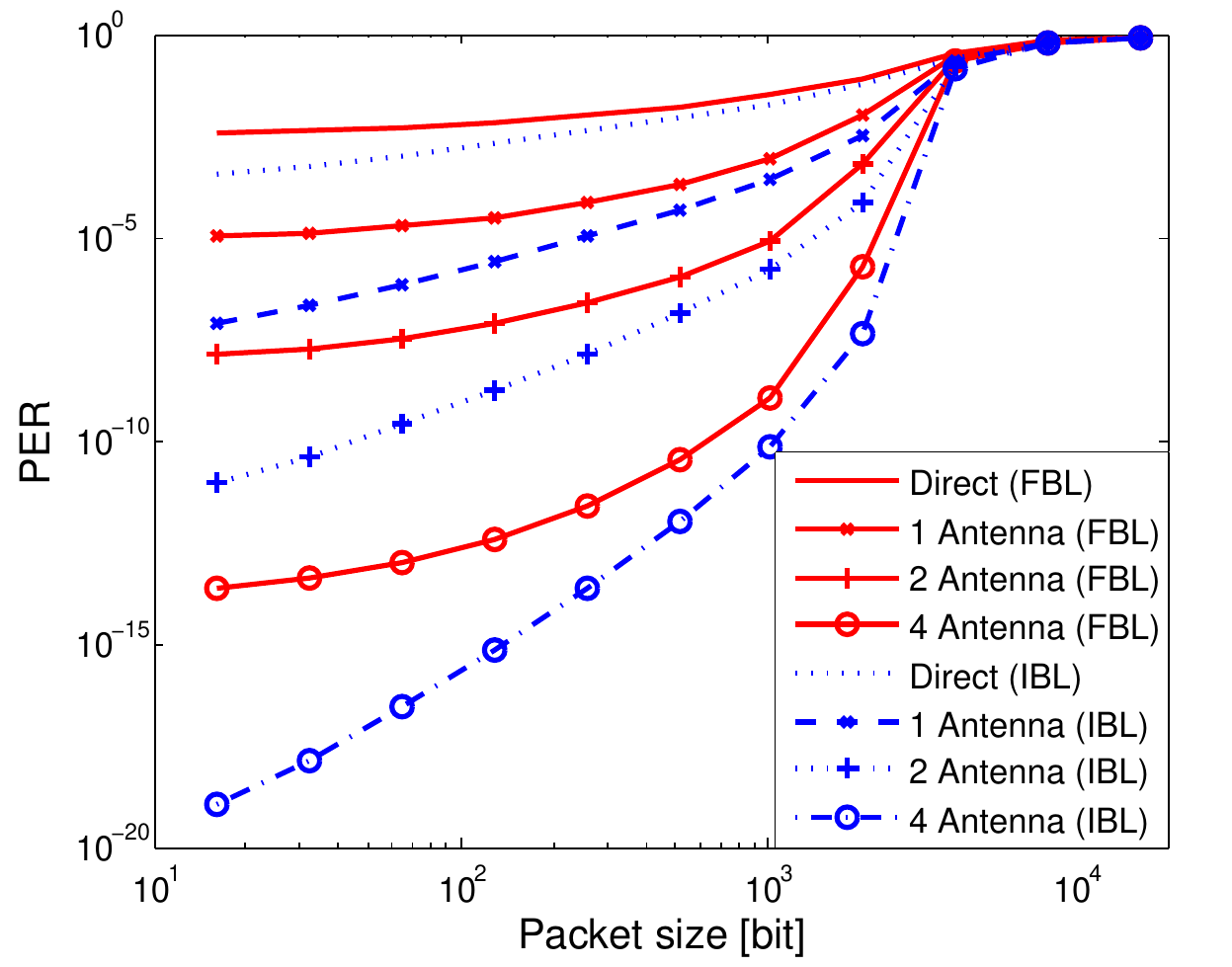}
{\dname{}} {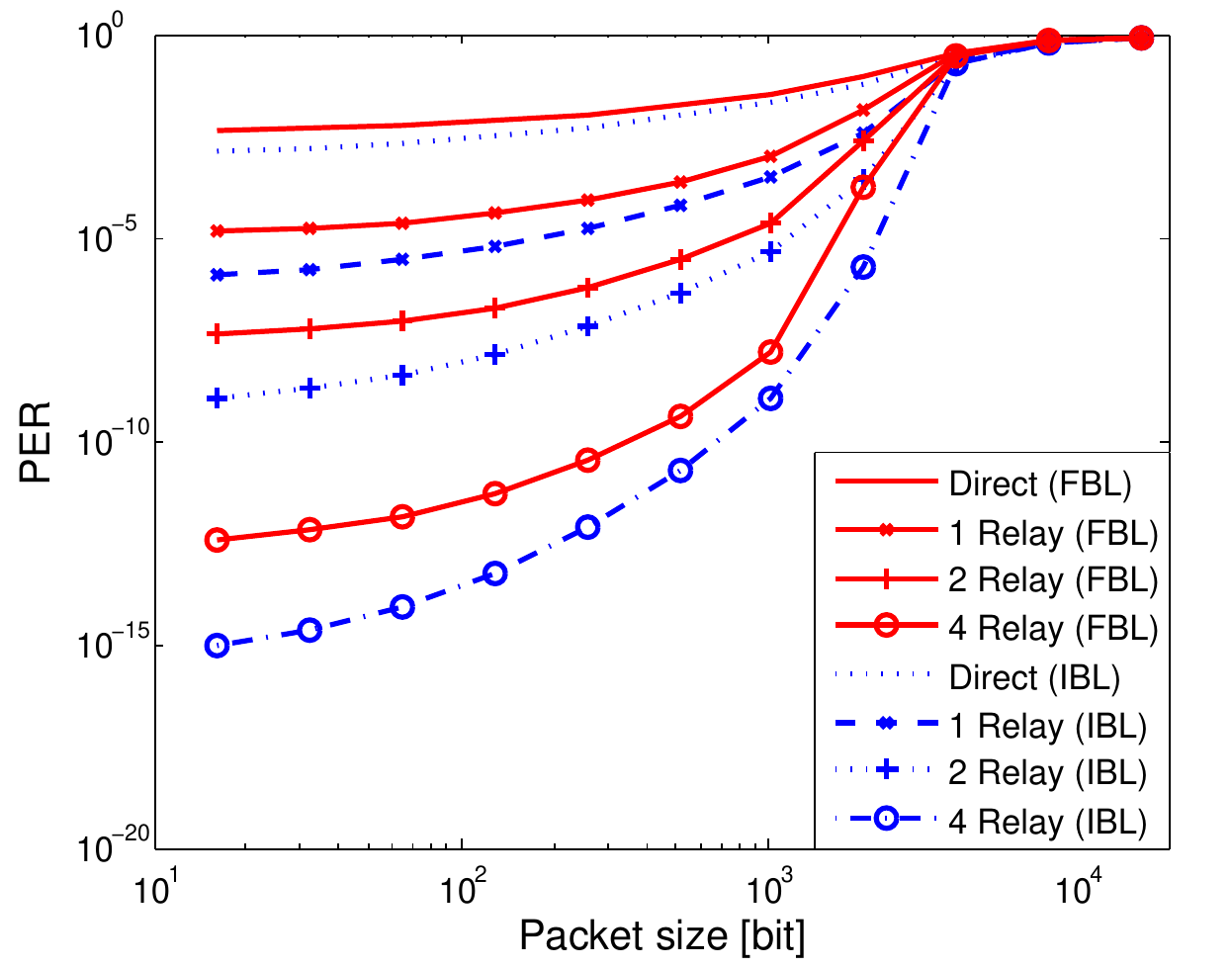}{t}{0.51}

We next are interested in the performance difference of the considered systems when utilizing either the \acp{FBL} or the \acp{IBL} modeling regime.
Therefore, we compare the \ac{PER} of \direct{}, \cname{}, and \dname{} under the \ac{IBL} and \ac{FBL} regime, varying different transmission parameters.
We begin with the packet size $D$, which we vary between \unit[$2^{4}$]{bit} and \unit[$2^{14}$]{bit}.
The results for \cname{} and \dname{} are depicted in \pref{fig:2regime_Packet}~(a) and~(b), respectively.

In general, a higher number of antennas or relays decreases the \ac{PER} due to an increasing cooperative diversity.
In addition, when approaching $D$ = \unit[$10^{4}$]{bit}, the \ac{PER} rapidly increases for both regimes as the available transmission symbols do not suffice to reliably transmit such large packets. 
More interestingly, for smaller packet sizes (below \unit[$10^{3}$]{bit}), we observe a significant gap (albeit in the logarithmic scaling) between system performance under the \ac{FBL} and the \ac{IBL} regime.
In the following, we provide an explanation for the observation while the rigorous proof will be considered in our future work.
Note that the fundamental difference between the \ac{FBL} and the \ac{IBL} regimes is that only the \ac{FBL} model considers decoding errors due to random noise. 
With smaller and smaller packets, the scheduling error due to fading decreases very much, which  allows us to set the target decoding error probability more aggressively, \ie much lower.
As in the figure we consider a fixed  target decoding error probability for different packet sizes, this makes the decoding error probability be dominant for the \ac{FBL} model when the packet size is small, in comparison to the scheduling error probability. 
Hence, improving the reliability by purely reducing the packet size is not quite efficient in the \ac{FBL} regime in comparison to the \ac{IBL} regime. 

In the \ac{IBL} regime, \cname{} clearly outperforms \dname{}, when the number of \ac{AP} antennas corresponds to the number of relays. 
Recall that in the relaying process of \cname{}, the \ac{AP} selects the best antenna for receiving a packet and, independently from the first choice, the best antenna for transmitting the packet.
This leads to a higher flexibility in the transmission path selection than in \dname{}, where the best (single-antenna) relay for receiving and transmitting is selected. 
Moreover, the overhead for acquiring instantaneous \ac{CSI} in \dname{} considerably increases with the number of potential relays and the number of terminals $N$, whereas in \cname{} the overhead only depends on $N$.
Nevertheless, the effects of \acp{FBL} dominate the \ac{PER} for smaller packets, such that the advantage of centrally relaying packets is lower than under the \ac{IBL} regime.

\renewcommand{\mycap}{Varying the \ac{SNR} $\overline{\gamma}$ for \cname{} and \dname{}.}
\doublefig{fig:2regime_SNR}
{\cname{}} {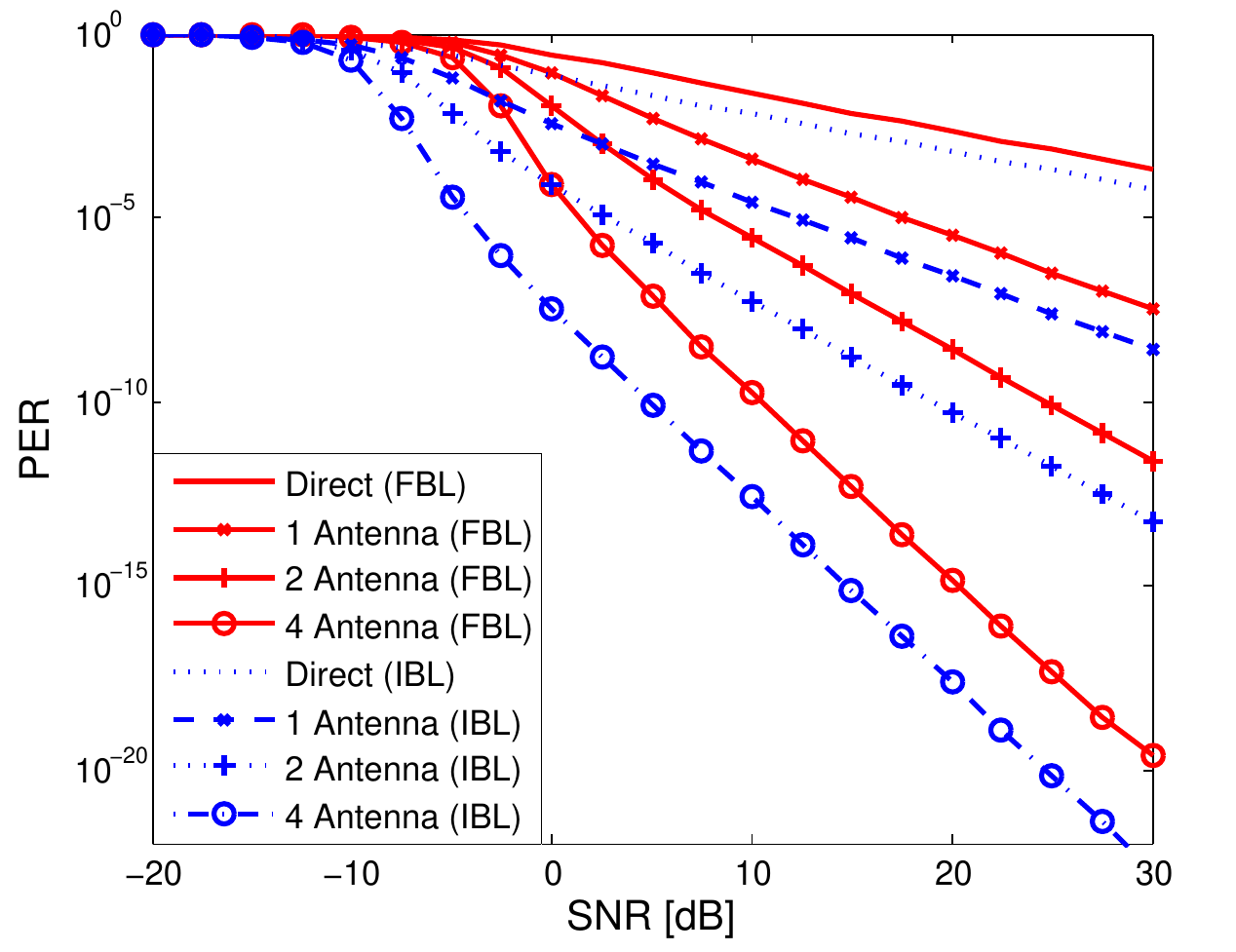}
{\dname{}} {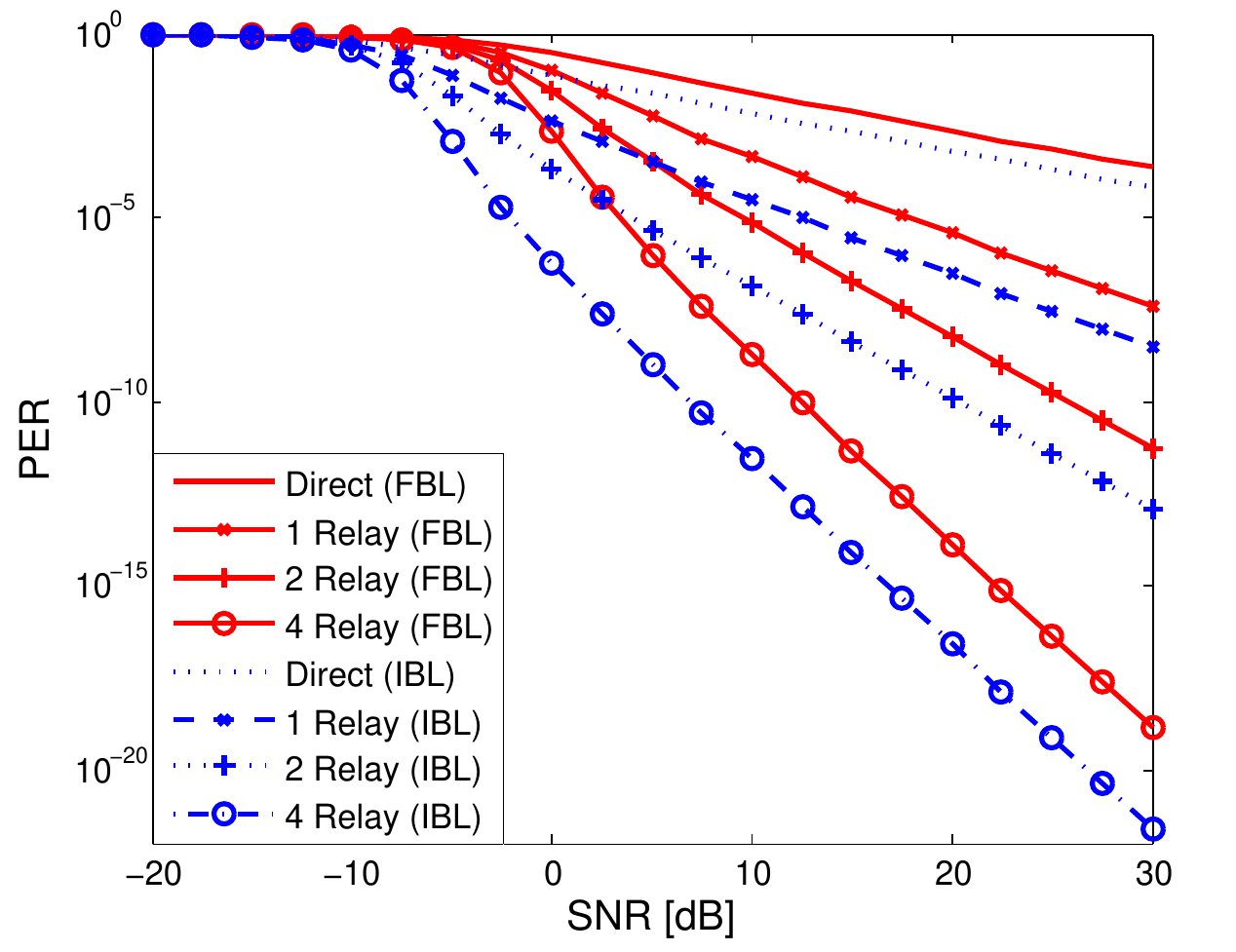}{t}{0.51}

Secondly, the relationship between \ac{PER} and average \ac{SNR} for the system variants \cname{} and \dname{} are shown in \pref{fig:2regime_SNR}.
In this scenario, the average receiver \ac{SNR} is varied (homogeneously for all links) from \unit[$-20$]{dB} to \unit[$30$]{dB}.
The aforementioned advantage of a higher flexibility in \cname{} becomes apparent in the \ac{PER} at $\gamma$ = \unit[$0$]{dB}.
Interestingly, for a fixed packet size $D$ the gap between \ac{FBL} and \ac{IBL} remains constant for a large range of \acp{SNR}. 
This indicates that in the high \ac{SNR} region the performance loss of reliability due to random noise error is not influenced by the SNR. 
In other words, improving the reliability by increasing the \ac{SNR} is efficient in both the \ac{FBL} regime  and the \ac{IBL} regime.
The figure finally reveals that with a moderate diversity degree (\ie three) a \ac{PER} of $10^{-10}$ should in principle be achievable already roughly from an average \ac{SNR} of \unit[$20$]{dB}, while an increase of the diversity degree to five reduces the required average \ac{SNR} down to \unit[$10$]{dB}.

\subsection{Scalability}

A central question of our work is how the performance of cooperative transmissions behaves with an increasing number of terminals when considering the overhead of collecting \ac{CSI} and the effects of finite blocklengths.
Recall that we assume that each terminal has one packet of size $D$ that must be transmitted within $T_\mathrm{cyc}$ = \unit[$1$]{ms}.
Thus, each additional terminal reduces statistically the available amount of symbols per transmission and increases the \ac{CSI} overhead.
In this context, our two relaying strategies, \cname{} and \dname{}, serve as a reference for two fundamental design decisions:
With central relaying the \ac{CSI} overhead only grows linearly with $N$ while the cooperative diversity is limited to the number of antennas at the \ac{AP}.
In turn, with decentralized relaying, the \ac{CSI} overhead grows quadratically in $N$ while the cooperative diversity increases with every additional terminal.

\renewcommand{\mycap}{Varying the number of transmissions/terminals $N$ for \cname{} and \dname{}.}
\doublefig{fig:2regime_N}
{\cname{}} {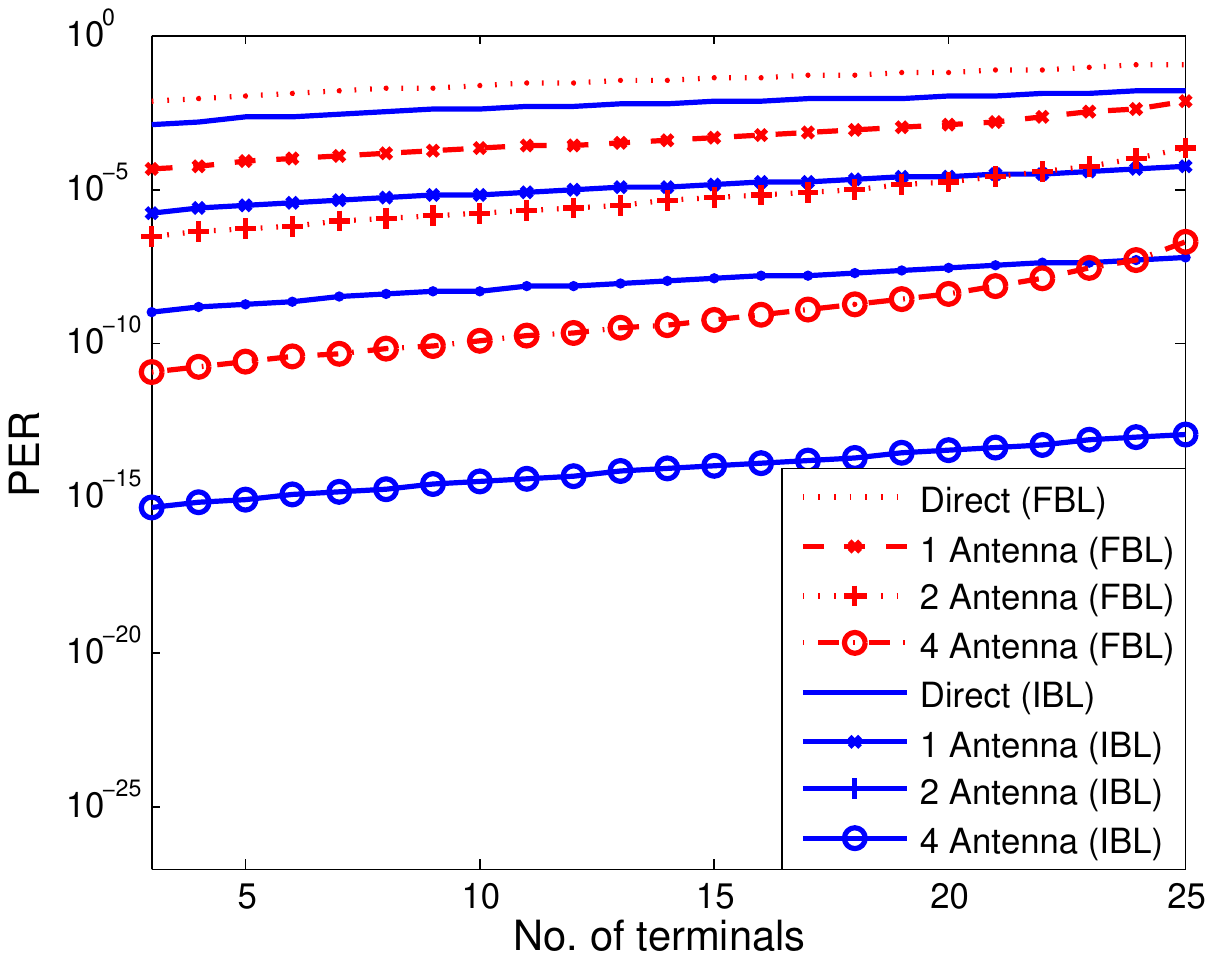}
{\dname{}} {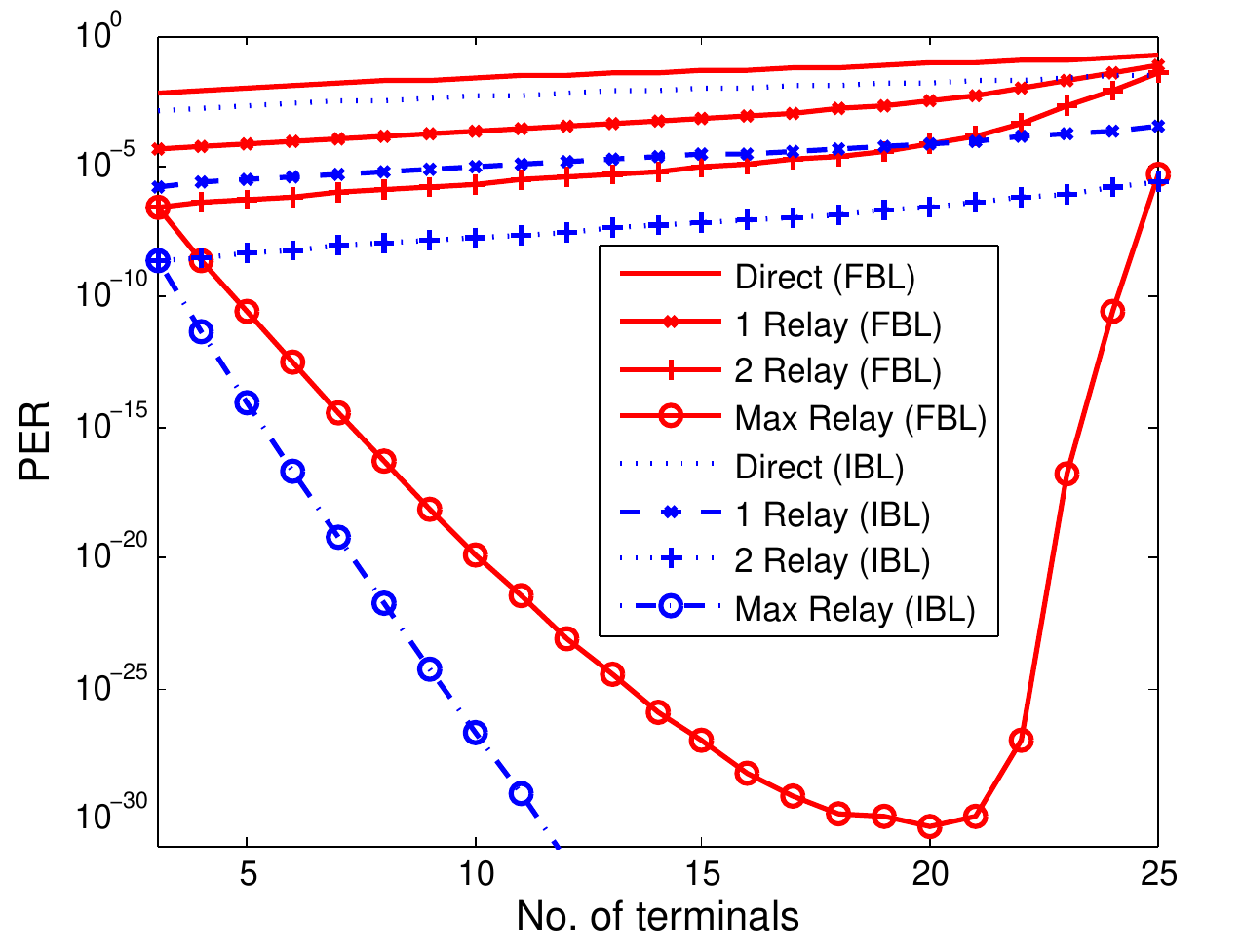}{t}{0.515}

In \pref{fig:2regime_N}, the \ac{PER} for \cname{}~(a) and \dname{}~(b) when increasing $N$ are shown.
Note that ``Max Relay'' in \dname{} denotes that all overhearing terminals, including the \ac{AP}, are considered as relay candidates. 
For \cname{}, each additional antenna at the \ac{AP} decreases the \ac{PER} by several orders of magnitude, as already seen before. 
In the \ac{IBL} regime, the achieved transmission reliability through cooperative diversity is almost insensitive to an increasing $N$.
In the \ac{FBL} regime, this is only true for the first part of the considered range.
At $N=20$, the slope of the \ac{PER} begins to change, emphasizing the additional impact of the decoding error which is present in the \acp{FBL} model.
Nevertheless, it can be stated that \cname{} has a relatively stable performance for the considered parametrization under both models.

For \dname{}, we observe a similar behavior as in \cname{} when the number of relays is limited. 
However, for the system set-up that utilizes the full diversity degree in the system, a significant performance improvement (\ie lower and lower PERs) can be observed with each additional terminal added to the system.
Note that this addition leads to a higher load as well as a higher overhead while on the other side the diversity order increases. 
The PER behavior is particularly visible for the results under the \ac{IBL} regime where the \ac{PER} decreases by two orders of magnitude with each additional terminal.
However, the results under the \ac{FBL} regime indicate that this behavior is not entirely accurate especially when many terminals are present in the system.
Although each terminal introduces additional cooperative diversity, the statistically effects of the reduced transmission symbols in combination with decoding error probability introduced by the \acp{FBL} model lead to a point of saturation where the reliability afterward drastically drops.
In practice, this saturation point can be shifted to the right by increasing the transmission resources or by limiting the \ac{CSI} overhead, \eg by locally dropping low-quality links instead of reporting every link to the \ac{AP}.

\renewcommand{\mycap}{Varying the overhead $(\alpha, \beta)$ and the channel bandwidth $B$ in \dname{} for an increasing $N$. Note that the values shown in the figure (especially these are below $10^{-30}$) are more of theoretical nature.}
\doublefig{fig:Varying_CSIcost}
{\acs{CSI} overhead $(\alpha, \beta)$.} {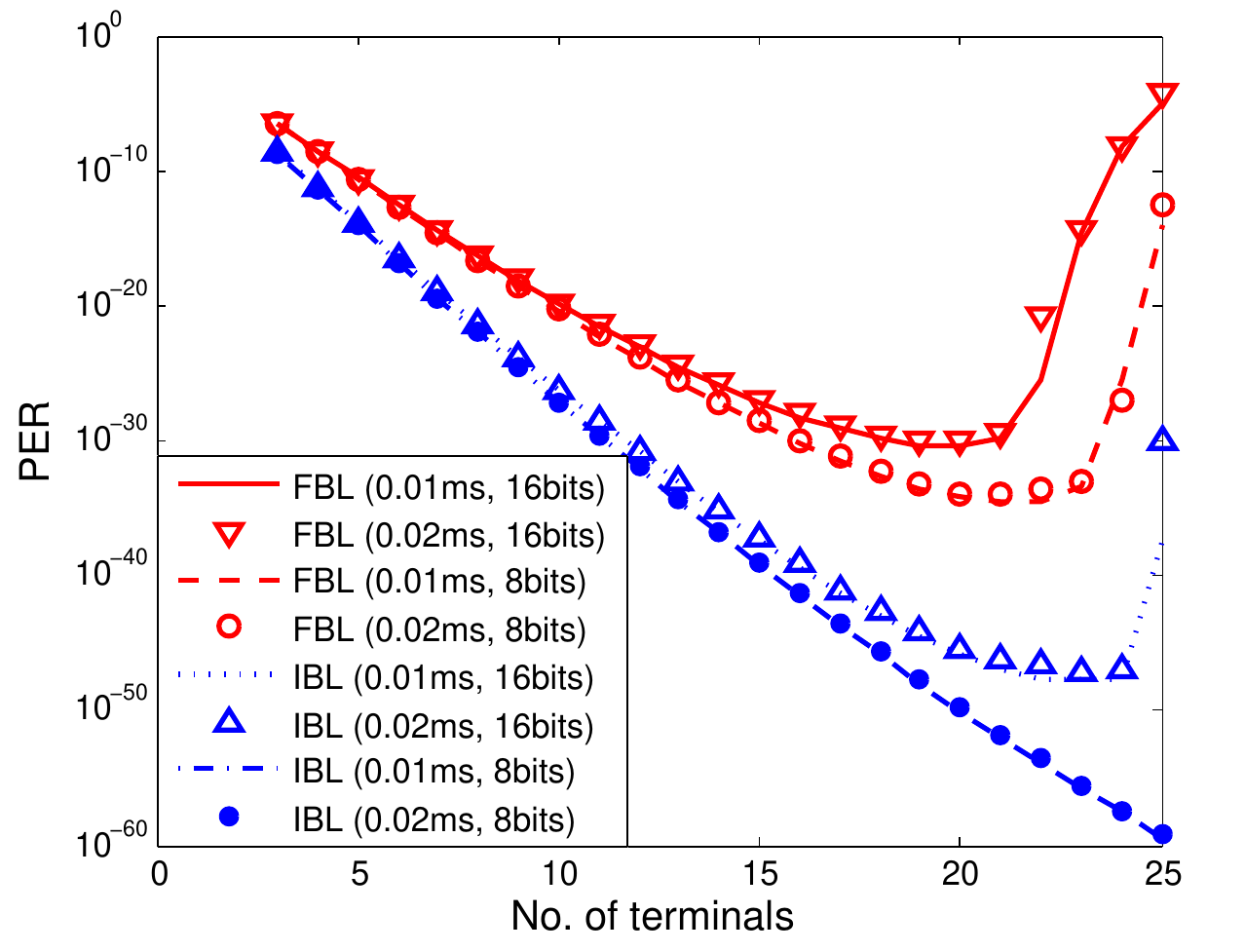}
{Channel bandwidth $B$.} {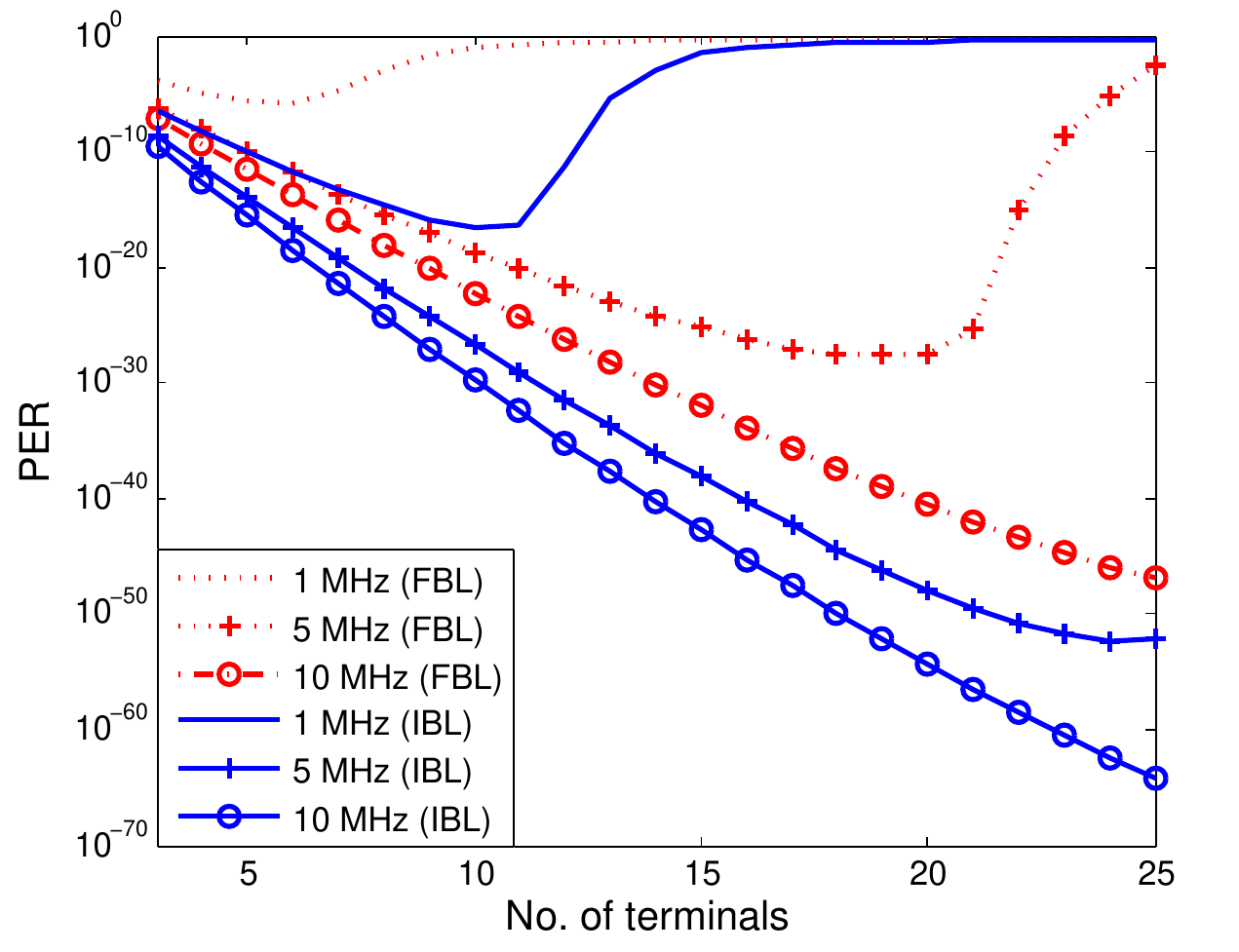}{t}{0.515}

In the following, we provide more details on the quasi-convex \ac{PER} when using all available relays.
In \pref{fig:Varying_CSIcost}~(a), we vary the overhead cost $(\alpha, \beta)$ to illustrate its impact on the system performance. 
For the \acp{IBL} and \acp{FBL} regime, doubling $\alpha$ does not significantly change the \ac{PER}.
In turn, when doubling $\beta$ the optimal \ac{PER} is higher and it is reached for a lower $N$.
Similarly, in \pref{fig:Varying_CSIcost}~(b)   the channel bandwidth $B$ is modified.
In this figure, the gap between \ac{IBL} and \ac{FBL} regime becomes even more visible.
According to our model under the \ac{IBL} regime, reliable communication at a small bandwidth $B$ = \unit[$1$]{MHz} is still feasible for $N=12$.
However, the \ac{FBL} results show that in this scenario a \ac{PER} below $10^{-9}$ is never reached.

\subsection{Target Error Probability}

\begin{figure}
	\centering
	\includegraphics[width=0.55\columnwidth]{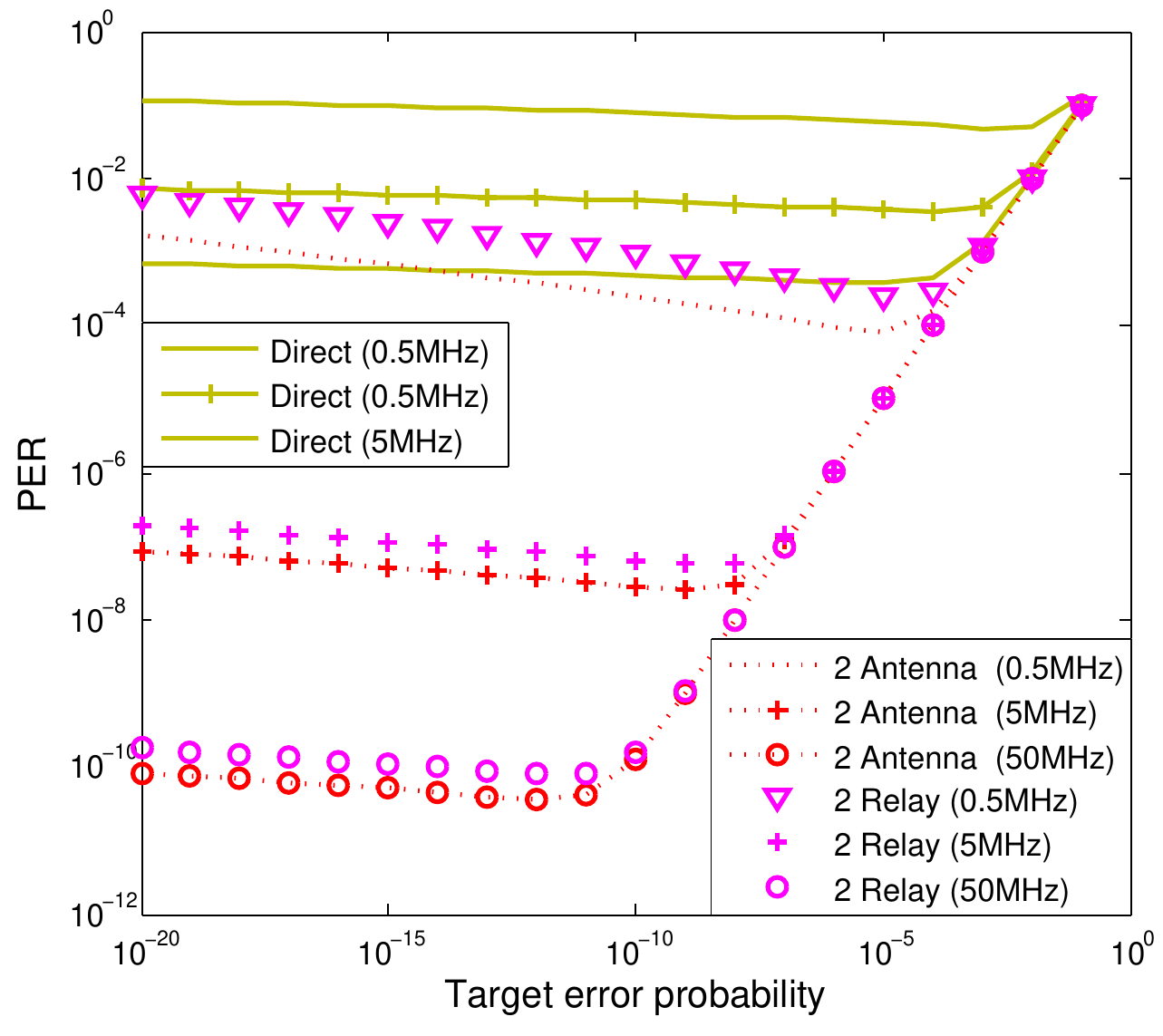}
	\caption{Under the finite blocklength regime with different bandwidth: PER vs. target error probability}
	\label{fig:Convex_target_error_VaringBW}
\end{figure}

In the last part of the evaluation, we come back to the target error probability under the \ac{FBL} regime.
Recall that in \pref{sec:validation}, we validated the convexity of the $\mathrm{PER}_\mathrm{FBL}$ in $\epsilon^*$.
It remains to show how the optimum is affected by the available transmission resources.
We thus additionally consider the scenarios of having few resources and having many resources, by setting the channel bandwidth $B$ to the corner cases of $B$ = \unit[$0.5$]{MHz} and $B$ = \unit[$50$]{MHz}, respectively.
The results for \cname{} and \dname{} with two available antennas/relays are shown in \pref{fig:Convex_target_error_VaringBW}. 

In all cases, the \ac{PER} curves are convex in $\epsilon^*$.
However, the slope on the left side of the optimum differs depending on available bandwidth and cooperative diversity.
For a narrow bandwidth ($B=\unit[0.5]{MHz}$), the slope of the \ac{PER} is steeper than for a wide bandwidth ($B$ = \unit[$50$]{MHz}).
Nevertheless, even for narrow bandwidths selecting a lower $\epsilon^*$ than the optimum results in a better system performance than selecting a higher one.
Hence, for practical systems where the optimal $\epsilon^*$  can not be determined, one should rather select a conservative decoding error probability $\epsilon$ as the penalty from the scheduling errors in terms of the PER is lower than the penalty from setting a too optimistic decoding error probability.

%% file: 08_conclusion.tex
\section{Conclusion}
\label{sec:conclusion}

In this work, we developed of a finite blocklength performance model for a multi-terminal wireless industrial network leveraging cooperative diversity. 
We studied two distinct relaying schemes with different degrees of diversity and the associated costs for acquiring instantaneous \ac{CSI} at the \ac{AP}.
We showed that under the \ac{FBL} regime the \ac{PER} of the studied network is convex in the target error probability of each link. 
We empirically validated our analytical models by simulation. 
Through numerical analysis, we found that \cname{} is in general more reliable than \dname{}, when the number of \ac{AP} antennas corresponds to the number of available relays. 
With a fixed number of antennas~/~relays, the PER increases with the number of associated terminals, as they are sharing the limited transmission resources. 
However, if in \dname{} each associated terminal is considered as a potential relay, the \ac{PER} is convex in the number of terminals due to the trade-off between additional cooperative diversity and increasing overhead for acquiring \ac{CSI}. 
Additionally, we showed the impact of the overhead $(\alpha, \beta)$ for acquiring \ac{CSI} on the system performance.
In particular, the evaluation results show that the communication overhead $\beta$ stronger influences the performance than the time overhead $\alpha$.
Finally, when choosing a target error probability $\epsilon^*$ we suggest to err on the lower target error probability side, as this will still lead to near-optimal performance.

%% file: 09_proof_2.tex
\section{Proof of Proposition 1}
According to~\pref{eq:PER_j}, regarding the \ac{PER} for a packet $j$, $j = 1,2,...N$, we have
\begin{IEEEeqnarray}{RCL}
\frac{{\partial {{\rm{PER}}_{{\rm{FBL}},j} }}}{{\partial {\varepsilon_{{\rm ave},j} ^*}}} & = &  - \frac{{\partial {p_j}}}{{\partial {\varepsilon_{{\rm ave},j}  ^*}}} + \frac{{\partial {p_j}}}{{\partial {\varepsilon_{{\rm ave},j}  ^*}}}{\varepsilon_{{\rm ave},j}  ^*} + {p_j} \quad , \nonumber \\
\frac{{{\partial ^2}{\rm{PE}}{{\rm{R}}_{{\rm{FBL}},j}}}}{{{\partial ^2}{\varepsilon_{{\rm ave},j}  ^*}}} & = &  - \frac{{{\partial ^2}{p_j}}}{{{\partial ^2}{\varepsilon_{{\rm ave},j}  ^*}}} + \frac{{{\partial ^2}{p_j}}}{{{\partial ^2}{\varepsilon_{{\rm ave},j}  ^*}}}{\varepsilon_{{\rm ave},j}  ^*} + 2\frac{{\partial {p_i}}}{{\partial {\varepsilon_{{\rm ave},j}  ^*}}} \nonumber \quad .
\end{IEEEeqnarray}

We first study the PER of packet~$1$ and subsequently, we will extend the analysis to packet $j$, with $j \ge 2$.
According to our system model, packet~$1$ could be transmitted either via the direct link or via the two-hop relaying. 
In the following, these two cases are discussed separately.
\begin{enumerate}
\item  If packet~$1$ is transmitted via the direct link, we have $ {\varepsilon_{{\rm ave},i}  ^*} ={\varepsilon  ^*}$.  
The probability of scheduling packet~$1$ is
${p_1} = \int\limits_{{\gamma ^*}/\bar \gamma }^{+\infty  } {{e^{-z}}dz}  = \frac{{{e^{ - {\gamma ^*}/\bar \gamma }}}}{{\bar \gamma }}$
with first and second derivatives with respect to $\varepsilon ^*$:
$\frac{{\partial {p_1}}}{{\partial {\varepsilon ^*}}} = - \frac{1}{{{{\bar \gamma }^2}}} \frac{{\partial {\gamma ^*}}}{{\partial {\varepsilon ^*}}}{e^{ - {\gamma ^*}/\bar \gamma }} $ and 
$\frac{{{\partial ^2}{p_1}}}{{{\partial ^2}{\varepsilon ^*}}} = \frac{1}{{{{\bar \gamma }^2}}}{e^{ - {\gamma ^*}/\bar \gamma }}\left( {\frac{1}{{\bar \gamma }}{{\left( {\frac{{\partial {\gamma ^*}}}{{\partial {\varepsilon ^*}}}} \right)}^2} - \frac{{{\partial ^2}{\gamma ^*}}}{{{\partial ^2}{\varepsilon ^*}}}} \right)$ . 

Therefore, we have:
\begin{IEEEeqnarray}{RCL}
\label{eq:_ref_}
\frac{{{\partial ^2}{\rm{PE}}{{\rm{R}}_{{\rm{FBL}},1}}}}{{{\partial ^2}{\varepsilon ^*}}} & = & 2\frac{{\partial {p_1}}}{{\partial {\varepsilon ^*}}} - \left( {1 - {\varepsilon ^*}} \right)\frac{{{\partial ^2}{p_1}}}{{{\partial ^2}{\varepsilon ^*}}} \nonumber \\
& = & \frac{1}{{{{\bar \gamma }^2}}}{e^{ - {\gamma ^*}/\bar \gamma }} \left\{ \left( 1 - {\varepsilon ^*} \right) \left( \frac{{{\partial ^2}{\gamma ^*}}}{\partial ^2}{\varepsilon ^*} - \frac{1}{\bar \gamma }{\left( \frac{{\partial {\gamma ^*}}}{{\partial {\varepsilon ^*}}} \right)^2} \right) - 2\frac{{\partial {\gamma ^*}}}{\partial {\varepsilon^*}} \right\} \quad .
\end{IEEEeqnarray}

Based on~\pref{eq:Polyanskiy_errorpro}, we have
\begin{IEEEeqnarray}{RCL}
{{\dot Q}^{ - 1}} \! \left( {{\varepsilon ^*}} \right) \! & = & \! \frac{{\sqrt M }}{{{{\log }_2}e}}\frac{{1 \! - \! \frac{1}{{\left( {{\gamma ^2} + 2\gamma } \right)}}\left( {{C_{{\rm{IBL}}}}(\gamma ) \! - \! D/M} \right)}}{{\sqrt {{\gamma ^2} + 2\gamma } }}\frac{{\partial {\gamma ^*}}}{{\partial {\varepsilon ^*}}} \quad . \nonumber
\end{IEEEeqnarray}

According to the definition of Q-function, the first derivative of ${{  Q}^{ - 1}}\left( {{\varepsilon ^*}} \right)$ with respect to $\varepsilon ^*$ is given by
\begin{IEEEeqnarray}{RCL}
{{\dot Q}^{ - 1}}\left( {{\varepsilon ^*}} \right) & = & - \sqrt {2\pi } {e^{\frac{{{{\left( {{Q^{ - 1}}\left( {{\varepsilon ^*}} \right)} \right)}^2}}}{2}}} < 0 \quad . \nonumber
\end{IEEEeqnarray}

Therefore, ${1 - \frac{1}{{\left( {{\gamma ^2} + 2\gamma } \right)}}\left( {{C_{{\rm{IBL}}}}(\gamma ) - D/M} \right)}>0$ as ${\gamma ^2} + 2\gamma  > {\log _2}\left( {1 + \gamma } \right) = {C_{{\rm{IBL}}}}(\gamma ) > {C_{{\rm{IBL}}}}(\gamma ) - D/M$ for $\gamma>0$.
Hence, ${\frac{{{\partial }{\gamma ^*}}}{{{\partial }{\varepsilon ^*}}}}<0$.
In  particular, we have
\begin{IEEEeqnarray}{RL}
\frac{{\bar \gamma }}{2}\frac{{\partial {\gamma ^*}}}{{\partial {\varepsilon ^*}}} & = \frac{{\bar \gamma }}{2}\frac{{ - \sqrt {2\pi } {e^{{{{\left( {{Q^{ - 1}}\left( {{\varepsilon^*}} \right)} \right)}^2}}/2}}}}{{\frac{{\sqrt M }}{{{{\log }_2}e}}\frac{{1 - \frac{1}{{\left( {{\gamma ^2} + 2\gamma } \right)}}\left( {{C_{{\rm{IBL}}}}(\gamma ) - D/M} \right)}}{{\sqrt {{\gamma ^2} + 2\gamma } }}}} 
 <  \!\! - \bar \gamma \sqrt {\frac{\left( {{\gamma ^2} + 2\gamma } \right)}{M}}  \cdot {e^{\!{M{{\left( {1 + \gamma } \right)}^2}{{\left( {\frac{{{C_{{\rm{IBL}}}}(\gamma ) - D/M}}{{{{\log }_2}e\sqrt {\left( {{\gamma ^2} + 2\gamma } \right)} }}} \right)}^2}}\!\!/2}}
\!\! \ll \!\! -1 \ . \nonumber
\end{IEEEeqnarray}

Similarly,  the second derivative of ${{  Q}^{ - 1}}\left( {{\varepsilon ^*}} \right)$ with respect to $\varepsilon ^*$ can be derived, based on~\pref{eq:Polyanskiy_errorpro} and the definition of Q-function, as
\begin{IEEEeqnarray}{RL}
{{\ddot Q}^{ - 1}}\left( {{\varepsilon ^*}} \right) = & \frac{{\sqrt M }}{{{{\log }_2}e}}\frac{{1 - \frac{1}{{\left( {{\gamma ^2} + 2\gamma } \right)}}\left( {{C_{{\rm{IBL}}}}(\gamma ) - \frac{D}{M}} \right)}}{{\sqrt {{\gamma ^2} + 2\gamma } }}\frac{{{\partial ^2}{\gamma ^*}}}{{{\partial ^2}{\varepsilon ^*}}} \\
& - \frac{{\sqrt M }}{{{{\log }_2}e}}\frac{{1 - \frac{1}{{\left( {{\gamma ^2} + 2\gamma } \right)}}\left( {{C_{{\rm{IBL}}}}(\gamma ) - \frac{D}{M}} \right)}}{{{{\left( {{\gamma ^2} + 2\gamma } \right)}^{\frac{3}{2}}}}}{{\left( {\frac{{\partial {\gamma ^*}}}{{\partial {\varepsilon ^*}}}} \right)}^2} \quad , \nonumber \\
{{\ddot Q}^{ - 1}}\left( {{\varepsilon ^*}} \right) & = 2\pi {Q^{ - 1}}\left( {{\varepsilon ^*}} \right){e^{{{\left( {{Q^{ - 1}}\left( {{\varepsilon ^*}} \right)} \right)}^2}}} > 0, \quad {\varepsilon ^*} < 0.5 \quad . \nonumber 
\end{IEEEeqnarray}

Moreover, we have ${\frac{{{\partial ^2}{\gamma ^*}}}{{{\partial ^2}{\varepsilon ^*}}}}<0$, then
\begin{IEEEeqnarray}{C}
{\frac{{{\partial ^2}{\rm{PE}}{{\rm{R}}_{{\rm{FBL}},1}}}}{{{\partial ^2}{\varepsilon ^*}}} > \frac{1}{{{{\bar \gamma }^3}}}{e^{ - {\gamma ^*}/\bar \gamma }}\frac{{\partial {\gamma ^*}}}{{\partial {\varepsilon ^*}}}\left( { - 2 - \bar \gamma \frac{{\partial {\gamma ^*}}}{{\partial {\varepsilon ^*}}}} \right)} > 0 \ , \nonumber
\end{IEEEeqnarray}
as $\frac{{\bar \gamma }}{2}\frac{{\partial {\gamma ^*}}}{{\partial {\varepsilon ^*}}}<-1$.
Hence, $\frac{{{\partial ^2}{\rm{PE}}{{\rm{R}}_{{\rm{FBL}},1}}}}{{{\partial ^2}{\varepsilon ^*}}} >0$ for the direct transmission case.

\item If packet $1$ is relayed via a two-hop link, we have $ {\varepsilon_{{\rm ave},i}  ^*} =2{\varepsilon  ^*} $. 
Then, the \ac{PER} of this packet is given by ${\rm{PE}}{{\rm{R}}_{{\rm{FBL}},1}} =   1 - {p_1} + 2{\varepsilon ^*}{p_1}$. Hence, the first and second derivatives of  the  \ac{PER} with respect to ${\varepsilon  ^*}$ are given by 
$\frac{{\partial {\rm{PE}}{{\rm{R}}_{{\rm{FBL}},1}}}}{{\partial {\varepsilon ^*}}} =  - \frac{{\partial {p_1}}}{{\partial {\varepsilon ^*}}}\left( {1 - 2{\varepsilon ^*}} \right) + 2{p_1}$ and
\small
\begin{IEEEeqnarray}{RL}
\frac{{{\partial ^2}{\rm{PE}}{{\rm{R}}_{{\rm{FBL}},1}}}}{{{\partial ^2}{\varepsilon ^*}}} & =  - \frac{{{\partial ^2}{p_1}}}{{{\partial ^2}{\varepsilon ^*}}}\left( {1 - 2{\varepsilon ^*}} \right) + \left( {{\varepsilon ^*} + 2} \right)\frac{{\partial {p_1}}}{{\partial {\varepsilon ^*}}} \nonumber \\
 = & - \! \left( {{\varepsilon ^*} + 2} \right)\frac{1}{{{{\bar \gamma }^2}}} \! \cdot \! \frac{{\partial {\gamma ^*}}}{{\partial {\varepsilon ^*}}}{e^{ - {\gamma ^*}/\bar \gamma }} \otimes {f_{{M_{R2}}}}(S) - \! \left( {1 \! - \! 2{\varepsilon ^*}} \right)\frac{1}{{{{\bar \gamma }^2}}}e^{\frac{{ - {\gamma ^*}}}{{\bar \gamma }}} \! \left( {\frac{1}{{\bar \gamma }}{{\left( {\frac{{\partial {\gamma ^*}}}{{\partial {\varepsilon ^*}}}} \right)}^2} \! - \! \frac{{{\partial ^2}{\gamma ^*}}}{{{\partial ^2}{\varepsilon ^*}}}} \right) \otimes {f_{{M_{R2}}}}(S) \nonumber \\
 = & \frac{1}{{{{\bar \gamma }^2}}}{e^{\frac{{ - {\gamma ^*}}}{{\bar \gamma }}}}\!\!\left\{  \!-\! \left( {{\varepsilon ^*} \!+\! 2} \right)\frac{{\partial {\gamma ^*}}}{{\partial {\varepsilon ^*}}} \!-\! ( {1 \!-\! 2{\varepsilon ^*}} )( \frac{1}{{\bar \gamma }}{{( {\frac{{\partial {\gamma ^*}}}{{\partial {\varepsilon ^*}}}} )}^2} \!\!\!-\! \frac{{{\partial ^2}{\gamma ^*}}}{{{\partial ^2}{\varepsilon ^*}}} ) \right\} 
  \otimes\! {f_{{M_{R2}}}}(S) \nonumber \\
 > & \frac{1}{{{{\bar \gamma }^2}}}{e^{\frac{{ \!-\! {\gamma ^*}}}{{\bar \gamma }}}}\!\!\left\{ { \!-\! 2\frac{{\partial {\gamma ^*}}}{{\partial {\varepsilon ^*}}} \!-\! \left( {1 \!-\! {\varepsilon ^*}} \right) ( {\frac{1}{{\bar \gamma }}{{\left( {\frac{{\partial {\gamma ^*}}}{{\partial {\varepsilon ^*}}}} \right)}^2} \!-\! \frac{{{\partial ^2}{\gamma ^*}}}{{{\partial ^2}{\varepsilon ^*}}}} )} \right\}  
  \otimes\! {f_{{M_{R2}}}}(S) > 0 \quad . \nonumber
\end{IEEEeqnarray}
\normalsize

Note that it has been shown in 1) that $\frac{{\partial {\gamma ^*}}}{{\partial {\varepsilon ^*}}}<0$ and in particular in~\pref{eq:_ref_} that 
\begin{IEEEeqnarray}{L}
 - 2\frac{{\partial {\gamma ^*}}}{{\partial {\varepsilon ^*}}} - \left( {1 - {\varepsilon ^*}} \right)\left( {\frac{1}{{\bar \gamma }}{{\left( {\frac{{\partial {\gamma ^*}}}{{\partial {\varepsilon ^*}}}} \right)}^2} - \frac{{{\partial ^2}{\gamma ^*}}}{{{\partial ^2}{\varepsilon ^*}}}} \right)>0 \quad , \nonumber  
\end{IEEEeqnarray}
thus we have  $\frac{{{\partial ^2}{\rm{PE}}{{\rm{R}}_{{\rm{FBL}},1}}}}{{{\partial ^2}{\varepsilon ^*}}} >0$ for the relaying case.

\end{enumerate}

So far, we have shown the convexity of the \ac{PER} of packet~$1$ with respect to ${\varepsilon ^*}$ for the direct transmission and the relaying case.
Note that due to random channel fading packet~$1$ is either transmitted directly or via a relay.
Hence, the expected \ac{PER} of packet~$1$ is the sum of the weighted \acp{PER} of these two cases, while the weights are probabilities with non-negative values.
Therefore, ${\rm{PER}}_{{\rm{FBL}},1}$ is convex in $\varepsilon ^*$.

Regarding the \ac{PER} of a packet~$j$, $j\ge 2$, we have, according to~\pref{eq:error_of_scheduling_each_j}, $\frac{{\partial {p_j}}}{{\partial {\varepsilon ^*}}} = \frac{{\partial {p_1}}}{{\partial {\varepsilon ^*}}} \otimes{f_{{M_{{\rm{min}},2}}}}(S)\otimes...\otimes {f_{{M_{{\rm{min}},j}}}}(S)$ and
\begin{IEEEeqnarray}{RL}
\frac{{{\partial ^2}{\rm{PE}}{{\rm{R}}_{{\rm{FBL}},2}}}}{{{\partial ^2}{\varepsilon ^*}}} & =  - \frac{{{\partial ^2}{p_2}}}{{{\partial ^2}{\varepsilon ^*}}} + \frac{{{\partial ^2}{p_2}}}{{{\partial ^2}{\varepsilon ^*}}}{\varepsilon ^*} + 2\frac{{\partial {p_2}}}{{\partial {\varepsilon ^*}}} \nonumber \\
 = & \! \left( {{\varepsilon ^*} \! - \! 1} \right)\frac{{{\partial ^2}{p_1}}}{{{\partial ^2}{\varepsilon ^*}}} \! \otimes \! {f_{{M_{{\rm{min}},2}}}}(S) \ldots \! \otimes \! {f_{{M_{{\rm{min}},j}}}}(S) \! + \! 2{f_{{M_{{\rm{min}},{\rm{1}}}}}}(S) \! \otimes \! {f_{{M_{{\rm{min}},2}}}}(S) \ldots \! \otimes \! {f_{{M_{{\rm{min}},j}}}}(S) \nonumber \\
 = & \left( {\frac{{{\partial ^2}{p_1}}}{{{\partial ^2}{\varepsilon ^*}}}\left( {{\varepsilon ^*} \!-\! 1} \right) \! +\!  2\frac{{\partial {p_1}}}{{\partial {\varepsilon ^*}}}} \right) \otimes {f_{{M_{{\rm{min}},2}}}}(S) \ldots  
  \otimes {f_{{M_{{\rm{min}},j}}}}(S) > 0 \quad . \nonumber
\end{IEEEeqnarray}

Hence, ${\rm{PER}}_{{\rm{FBL}},j}$ is  convex in $\varepsilon ^*$ for $j = 1,2,...,N$.
As the sum of convex functions is also convex,  ${\rm{PER}}_{{\rm{FBL}}}  = \frac{1}{N}\sum\limits_{j = 1}^N {\rm{PER}}_{{\rm{FBL}},j}  $ is convex in  $\varepsilon ^*$.

%% file: journal.bbl
\begin{thebibliography}{10}
\providecommand{\url}[1]{#1}
\csname url@samestyle\endcsname
\providecommand{\newblock}{\relax}
\providecommand{\bibinfo}[2]{#2}
\providecommand{\BIBentrySTDinterwordspacing}{\spaceskip=0pt\relax}
\providecommand{\BIBentryALTinterwordstretchfactor}{4}
\providecommand{\BIBentryALTinterwordspacing}{\spaceskip=\fontdimen2\font plus
\BIBentryALTinterwordstretchfactor\fontdimen3\font minus
  \fontdimen4\font\relax}
\providecommand{\BIBforeignlanguage}[2]{{%
\expandafter\ifx\csname l@#1\endcsname\relax
\typeout{** WARNING: IEEEtran.bst: No hyphenation pattern has been}%
\typeout{** loaded for the language `#1'. Using the pattern for}%
\typeout{** the default language instead.}%
\else
\language=\csname l@#1\endcsname
\fi
#2}}
\providecommand{\BIBdecl}{\relax}
\BIBdecl

\bibitem{ABC+14}
J.~G. Andrews, S.~Buzzi, W.~Choi, S.~V. Hanly, A.~Lozano, A.~C.~K. Soong, and
  J.~C. Zhang, ``{What Will 5G Be?}'' \emph{IEEE Journal on Selected Areas in
  Communications}, vol.~32, no.~6, pp. 1065--1082, Jun. 2014.

\bibitem{FWB+14}
A.~Frotzscher, U.~Wetzker, M.~Bauer \emph{et~al.}, ``{Requirements and Current
  Solutions of Wireless Communication in Industrial Automation},'' in
  \emph{IEEE Int'l Conf. on Comm. Workshops (ICC)}, Jun. 2014, pp. 67--72.

\bibitem{Neum07}
P.~Neumann, ``{Communication in Industrial Automation--What Is Going On?}''
  \emph{Control Engin. Practice}, vol.~15, no.~11, pp. 1332--1347, 2007.

\bibitem{DASC04}
S.~N. Diggavi, N.~Al-Dhahir, A.~Stamoulis, and A.~R. Calderbank, ``{Great
  Expectations: The Value of Spatial Diversity in Wireless Networks},''
  \emph{Proceedings of the IEEE}, vol.~92, no.~2, pp. 219--270, Feb. 2004.

\bibitem{LTWo04}
J.~N. Laneman, D.~N.~C. Tse, and G.~W. Wornell, ``{Cooperative Diversity in
  Wireless Networks: Efficient Protocols and Outage Behavior},'' \emph{IEEE
  Trans. on Inform. Theory}, vol.~50, no.~12, pp. 3062--3080, Dec. 2004.

\bibitem{BKRL06}
A.~Bletsas, A.~Khisti, D.~P. Reed, and A.~Lippman, ``{A Simple Cooperative
  Diversity Method Based on Network Path Selection},'' \emph{IEEE Journal on
  Selected Areas in Comm.}, vol.~24, no.~3, pp. 659--672, Mar. 2006.

\bibitem{BlLi06}
A.~Bletsas and A.~Lippman, ``{Implementing Cooperative Diversity Antenna Arrays
  with Commodity Hardware},'' \emph{IEEE Communications Magazine}, vol.~44,
  no.~12, pp. 33--40, Dec. 2006.

\bibitem{IkAh10}
S.~S. Ikki and M.~H. Ahmed, ``{Performance Analysis of Adaptive
  Decode-and-Forward Cooperative Diversity Networks with Best-Relay
  Selection},'' \emph{IEEE Trans. on Comm.}, vol.~58, no.~1, pp. 68--72, Jan.
  2010.

\bibitem{ChSe16}
A.~Chaaban and A.~Sezgin, ``{Multi-Hop Relaying: An End-to-End Delay
  Analysis},'' \emph{IEEE Transactions on Wireless Communications}, vol.~15,
  no.~4, pp. 2552--2561, Apr. 2016.

\bibitem{SSR+15}
V.~N. Swamy, S.~Suri, P.~Rigge \emph{et~al.}, ``{Cooperative Communication for
  High-Reliability Low-Latency Wireless Control},'' in \emph{IEEE Int'l Conf.
  on Comm. (ICC)}, Jun. 2015, pp. 4380--4386.

\bibitem{DoGr15}
C.~Dombrowski and J.~Gross, ``{EchoRing: A Low-Latency, Reliable Token-Passing
  MAC Protocol for Wireless Industrial Networks},'' in \emph{Proc. of 21th
  European Wireless Conference (EW15)}, May 2015, pp. 1--8.

\bibitem{SDWG15}
M.~Serror, C.~Dombrowski, K.~Wehrle, and J.~Gross, ``{Channel Coding Versus
  Cooperative ARQ: Reducing Outage Probability in Ultra-Low Latency Wireless
  Communications},'' in \emph{{IEEE Global Comm. Conf. (GLOBECOM) Workshops
  (ULTRA$^2$)}}, San Diego, USA, Dec. 2015.

\bibitem{Verdu_2010}
Y.~Polyanskiy, H.~Poor, and S.~Verdu, ``{Channel Coding Rate in the Finite
  Blocklength Regime},'' \emph{IEEE Trans. on Inform. Theory}, vol.~56, no.~5,
  pp. 2307--2359, 2010.

\bibitem{SHD+16}
M.~Serror, Y.~Hu, C.~Dombrowski, K.~Wehrle, and J.~Gross, ``{Performance
  Analysis of Cooperative {ARQ} Systems for Wireless Industrial Networks},'' in
  \emph{IEEE Int'l Symp. on a World of Wireless, Mobile and Multim. Netw. (IEEE
  WoWMoM 2016)}, Coimbra, Portugal, Jun. 2016.

\bibitem{Hu_2015}
Y.~Hu, J.~Gross, and A.~Schmeink, ``{On the Capacity of Relaying with Finite
  Blocklength},'' \emph{IEEE Trans. Veh. Technol.}, vol.~62, no.~5, pp.
  1490--1502, Mar. 2015.

\bibitem{Hu_letter_2015}
------, ``{On the Performance Advantage of Relaying Under the Finite
  Blocklength Regime},'' \emph{IEEE Comm. Letter}, vol.~62, no.~5, pp.
  1490--1502, Jul. 2015.

\bibitem{Hu_TWC_2016}
------, ``{Blocklength-limited Performance of Relaying Under Quasi-Static
  Rayleigh Channels},'' \emph{IEEE Trans. on Wireless Comm.}, accepted.

\bibitem{Yang_2014}
Y.~Wei, G.~Durisi, T.~Koch, and Y.~Polyanskiy, ``{Quasi-Static Multiple-Antenna
  Fading Channels at Finite Blocklength},'' \emph{IEEE Transactions on
  Information Theory}, vol.~60, no.~7, pp. 4232--4265, Jul. 2014.

\bibitem{Gursoy_2013}
M.~C. Gursoy, ``{Throughput Analysis of Buffer-constrained Wireless Systems in
  the Finite Blocklength Regime},'' \emph{EURASIP Journal on Wireless
  Communications and Networking}, vol. 2013, no.~1, 2013.

\bibitem{Peng_2011}
P.~Wu and N.~Jindal, ``{Coding Versus ARQ in Fading Channels: How Reliable
  Should the PHY Be?}'' \emph{IEEE Transactions on Communications}, vol.~59,
  no.~12, pp. 3363--3374, Dec. 2011.

\bibitem{Makki_2014}
B.~Makki, T.~Svensson, and M.~Zorzi, ``{Finite Block-Length Analysis of the
  Incremental Redundancy HARQ},'' \emph{IEEE Wireless Comm. Letters}, vol.~3,
  no.~5, pp. 529--532, Oct. 2014.

\bibitem{Makki_2015}
------, ``{Finite Block-Length Analysis of Spectrum Sharing Networks Using Rate
  Adaptation},'' \emph{IEEE Transactions on Communications}, vol.~63, no.~8,
  pp. 2823--2835, Aug. 2015.

\bibitem{Polyanskiy_2011}
Y.~Polyanskiy, H.~V. Poor, and S.~Verdu, ``{Dispersion of the Gilbert-Elliott
  Channel},'' \emph{IEEE Transactions on Information Theory}, vol.~57, no.~4,
  pp. 1829--1848, Apr. 2011.

\end{thebibliography}
